\documentstyle{mn}
 
\newif\ifAMStwofonts 
 
\newcommand{\mxum} {\bf {m1} \rm}
\newcommand{\mxdois} {\bf {m2} \rm}
\newcommand{\mxtres} {\bf {m3} \rm}

\ifoldfss 
  \ifCUPmtlplainloaded \else 
    \NewTextAlphabet{textbfit} {cmbxti10} {} 
    \NewTextAlphabet{textbfss} {cmssbx10} {} 
    \NewMathAlphabet{mathbfit} {cmbxti10} {} 
    \NewMathAlphabet{mathbfss} {cmssbx10} {} 
  \fi 
  \ifAMStwofonts 
    \ifCUPmtlplainloaded \else 
      \NewSymbolFont{upmath} {eurm10} 
      \NewSymbolFont{AMSa} {msam10} 
      \NewMathSymbol{\upi}     {0}{upmath}{19} 
      \NewMathSymbol{\umu}     {0}{upmath}{16} 
      \NewMathSymbol{\upartial}{0}{upmath}{40} 
      \NewMathSymbol{\leqslant}{3}{AMSa}{36} 
      \NewMathSymbol{\geqslant}{3}{AMSa}{3E}

    \fi 
  \fi 
\fi 
 
\ifnfssone 
  \newmathalphabet{\mathit} 
  \addtoversion{normal}{\mathit}{cmr}{m}{it} 
  \addtoversion{bold}{\mathit}{cmr}{bx}{it} 
  \newmathalphabet{\mathbfit} 
  \addtoversion{normal}{\mathbfit}{cmr}{bx}{it} 
  \addtoversion{bold}{\mathbfit}{cmr}{bx}{it} 
  \newmathalphabet{\mathbfss} 
  \addtoversion{normal}{\mathbfss}{cmss}{bx}{n} 
  \addtoversion{bold}{\mathbfss}{cmss}{bx}{n} 
  \ifAMStwofonts 
    \ifCUPmtlplainloaded \else 
      \UseAMStwoboldmath 
      \makeatletter 
      \new@mathgroup\upmath@group 
      \define@mathgroup\mv@normal\upmath@group{eur}{m}{n} 
      \define@mathgroup\mv@bold\upmath@group{eur}{b}{n} 
      \edef\UPM{\hexnumber\upmath@group} 
      \new@mathgroup\amsa@group 
      \define@mathgroup\mv@normal\amsa@group{msa}{m}{n} 
      \define@mathgroup\mv@bold\amsa@group{msa}{m}{n} 
      \edef\AMSa{\hexnumber\amsa@group} 
      \makeatother 
      \mathchardef\upi="0\UPM19 
      \mathchardef\umu="0\UPM16 
      \mathchardef\upartial="0\UPM40 
      \mathchardef\leqslant="3\AMSa36 
      \mathchardef\geqslant="3\AMSa3E 
    \fi 
  \fi 
\fi 
 
\ifnfsstwo 
  \DeclareMathAlphabet{\mathbfit}{OT1}{cmr}{bx}{it} 
  \SetMathAlphabet\mathbfit{bold}{OT1}{cmr}{bx}{it} 
  \DeclareMathAlphabet{\mathbfss}{OT1}{cmss}{bx}{n} 
  \SetMathAlphabet\mathbfss{bold}{OT1}{cmss}{bx}{n} 
  \ifAMStwofonts 
    \ifCUPmtlplainloaded \else 
      \DeclareSymbolFont{UPM}{U}{eur}{m}{n} 
      \SetSymbolFont{UPM}{bold}{U}{eur}{b}{n} 
      \DeclareSymbolFont{AMSa}{U}{msa}{m}{n} 
      \DeclareMathSymbol{\upi}{0}{UPM}{"19} 
      \DeclareMathSymbol{\umu}{0}{UPM}{"16} 
      \DeclareMathSymbol{\upartial}{0}{UPM}{"40} 
      \DeclareMathSymbol{\leqslant}{3}{AMSa}{"36} 
      \DeclareMathSymbol{\geqslant}{3}{AMSa}{"3E} 
    \fi 
  \fi 
\fi 
 
\ifCUPmtlplainloaded \else 
  \ifAMStwofonts \else 
    \def\upi{\pi} 
    \def\umu{\mu} 
    \def\upartial{\partial} 
  \fi 
\fi 
 
\title{Formation of rings in galactic discs by infalling small companions}
\author[E. Athanassoula, I. Puerari, A. Bosma]
       {E. Athanassoula, I. Puerari and A. Bosma\\ 
       Observatoire de Marseille\\
       2 Place Le Verrier\\
       F-13248 Marseille Cedex 4, France}
\date{Accepted .
      Received ;
      in original form 1995 November}

\pagerange{\pageref{firstpage}--\pageref{lastpage}}
\pubyear{1995}

\begin{document}

\maketitle

\label{firstpage}

\begin{abstract}
We use N-body simulations to study the formation of rings in a disc 
galaxy by the impact of a small spherical companion. Both barred and 
nonbarred target discs are considered. We discuss the effect of the 
properties of the target disc (distribution of mass in the disc, 
velocity dispersion, etc.) as well as of the mass and orbit of 
the companion on the properties of the rings, such as their amplitude, 
width, shape, expansion velocity and lifetime. In particular the amplitude, 
width, lifetime and expansion velocity of the first ring increase 
considerably with companion mass, and so does the expansion velocity 
of the particles in it and the total extent of the disc after the interaction. 
We also discuss the 
formation and evolution of spokes and show that they can be caused 
by companions of relatively small mass.
In our three examples of oblique impacts on barred target galaxies
we note important transient displacements of the bar, as well as changes 
of its pattern speed and size. An asymmetric 
pseudoring is formed in each case, and during the first stages of 
its evolution the bar forms part of it.

\end{abstract}


\section{Introduction}

	Ring galaxies show a pronounced ring structure surrounding an
apparently empty region in which an offcentered nucleus can often be seen. Such
objects are relatively rare and are mainly found in low density environments. 
Their properties have been reviewed and compared to those of disc ga\-la\-xies with 
resonant
rings (ringed galaxies) by Athanassoula \& Bosma (1985). Theys \& Spiegel
(1976) made the important remark that they have a companion which lies 
preferentially near
the minor axis of the ring, and this 
guided their simulations (Theys \&
Spiegel 1977) of a companion colliding with a disc galaxy, which indeed 
resulted in
the formation of rings in the disc.

	A clear picture of what happens during the collision is given by 
Lynds \& Toomre (1976) and by Toomre (1978). As the intruder approaches the 
disc, the extra inwards gravitational force it exerts on the disc particles 
increases and causes 
their orbits to contract. When the companion leaves there is a 
strong rebound. As a result the orbits crowd together and a high amplitude, 
transient density wave is formed, which propagates outwards. A second or third 
rebound is possible, resulting in a second or third ring.

	Self-consistent simulations following these precepts have been made 
recently by Huang \& Stewart (1988) and Appleton \& James (1990), while 
Hernquist \& Weil (1993) and 
Horelou \& Combes (1993) include also gas in the 
simulations (the latter, 
however, with a rigid companion and halo). Mihos \& 
Hernquist (1994) add star formation as well.
Relatively high mass companions, equal 
to 1, 0.4 or 0.333 times the target galaxy mass, are considered by Appleton \&
James (1990), Huang \& Stewart (1988) and Horelou \& Combes (1993),
respectively. Hernquist \& Weil (1993) use a compa\-nion mass equal to 
that of the target disc,
and about 0.25 of the total mass of the target,
for their fiducial simulation, but also present  
simulations with the companion having double or a quarter of that mass.

	In this paper we will use numerical simulations to investigate
further the formation of rings by infall of a small companion galaxy on a 
nonbarred or barred target galaxy. Prompted by the observational results for 
the Cartwheel galaxy (Davies \& Morton 1982, and discussion in Struck-Marcell 
\& Higdon 1993 and Appleton \& Struck-Marcell 1996), we use 
companions with comparatively low mass, i.e., 
0.02 to 0.2 times the target galaxy 
mass. This ensures that the disc survives
the collision and that the ring can be considered as a perturbation,
comparable to that of resonant rings of ringed galaxies. In fact one of our 
goals is to study the
morphology and kinematics of the rings 
formed by impacts and to compare them 
with the
corresponding ones for ringed galaxies. Moreover, we will consider 
the evolution in both 
nonbarred and barred target discs. 
In section \ref{sec:simul} we present
the initial conditions and numerical simulations and give some detail
on the computing techniques used. The results for the
impacts on nonbarred discs are presented in section \ref{sec:nonbar}. In
section \ref{sec:barred}  we discuss cases where the target disc is 
barred. A discussion about ring {\it versus} ringed
galaxies is given in section \ref{sec:ringringed}
and a summary of our results in section \ref{sec:summary}.

\section{Numerical techniques }
\label{sec:simul}

\subsection{Initial conditions }
\indent

The simulations 
we will discuss in this paper form part of a series which studies
the effects of a small companion on a disc galaxy. More information on
the setup of these simulations and their
evolution will be given elsewhere (Puerari \& Athanassoula,
in preparation). Here we will discuss only some characteristics
relevant to the problem of ring formation. 
The effects on the vertical structure of 
the disc will be left for a more global paper, encompassing a larger 
number of simulations.

Our model galaxies are
composed of a halo, a disc and, in some cases, a bulge, and some relevant
parameters of these components are given in Table~\ref{tab:models}. 
Four of
our models ({\bf m1}, {\bf m2}, {\bf m3} and {\bf s1}) have discs which were
initially setup with a truncated 
Kuzmin/Toomre (hereafter KT) projected radial density profile \\

${\displaystyle \Sigma (r) = \frac {M_D}{2\pi b_D^2}(1+\frac {r^2}{b_D^2})^{-3/2}}$ \\

\noindent (Kuzmin 1956, Toomre 1963) and a $\rm{sech}^2$({$z$}/{$z_0$}) 
vertical distribution, while their
halo and bulge have a truncated Plummer (hereafter PL) profile \\

${\displaystyle \rho (r) = \frac {3M}{4\pi b^3} (1+\frac {r^2}{b^2})^{-5/2}}$ \\

\noindent where $M$ and $b$ are the mass and scalelength of the component (halo, bulge, or companion). The radial
velocity dispersions in the disc are chosen so that the $Q$ parameter 
(Toomre 1964) \\

${\displaystyle Q = \frac {\sigma_R \kappa} {3.36 G \Sigma}}$ \\

\noindent is independent of radius and equal to 1, 1.1, 1.2 or 1.5. 
In the above formula 
$\sigma_R$ is the radial velocity dispersion, 
$\kappa$ is the epicyclic frequency,
$G$ is the gravitational constant and 
$\Sigma$ is the disc surface density.
We have calculated the
tangential velocity dispersion using the epicyclic approximation; the
vertical velocity dispersion follows from
the expression $\sigma^2_z=\pi G \Sigma z_0$ (Binney \& 
Tremaine 1987). The first section of Table~\ref{tab:models} describes 
the disc. Its first column gives the label of
the model, the second one the type of disc used (which for the four models we
are discussing is KT), the third gives the number of particles in the
disc, the fourth its mass, the fifth its scale length, the sixth its outer
cutoff radius, and the seventh its vertical scaleheight. 
The second and third section of Table~\ref{tab:models} give the 
same information,
but now for the halo and bulge respectively. For brevity we will often refer to model {\bf m1} as standard, and to models {\bf m2} and {\bf m3} as extended, or very extended.

\begin{table}
\centering
\caption{Model parameters}
\label{tab:models}
\begin{tabular}{@{}llrllrl@{}}
&&&&&& \\
\multicolumn{2}{l}{\bf DISC} &&&&& \\
&&&&&& \\
Model & Type & $N_D$ & $M_D$ & $b_D$ & $R_{D}$ & $z_0$ \\
{\bf m1} & KT & 8000 & 0.4 & 1.0 & 5.0 & 0.2 \\
{\bf m2} & KT & 8000 & 0.4 & 2.0 & 8.0 & 0.2 \\
{\bf m3} & KT & 8000 & 0.4 & 5.0 & 8.0 & 0.2 \\
{\bf s1} & KT & 8000 & 0.3215 & 1.0 & 5.0 & 0.2 \\
{\bf mb} & KT & 14000 & 0.7 & 1.0 & 5.0 & 0.2 \\
 & barred & 14000 & 0.7 & 0.84 & 5.0 & 0.21 \\
{\bf mh} & KT & 14000 & 0.7 & 1.0 & 5.0 & 0.2 \\
 & hot & 14000 & 0.7 & 0.83 & 5.0 & 0.21 \\
&&&&&& \\
\multicolumn{2}{l}{\bf HALO} &&&&& \\
&&&&&& \\
Model & Type & $N_H$ & $M_H$ & $b_H$ & $R_{H}$ & \\
{\bf m1} & PL & 30000 & 1.5 & 5.0 & 10.0 & \\
{\bf m2} & PL & 30000 & 1.5 & 5.0 & 10.0 & \\
{\bf m3} & PL & 30000 & 1.5 & 5.0 & 10.0 & \\
{\bf s1} & PL & 30000 & 1.0733 & 5.0 & 10.0 & \\
{\bf mb} & PL & 26000 & 1.3 & 10.0 & 10.0 & \\
   & PL & 26000 & 1.3  & --- & 11.8 & \\
{\bf mh} & PL & 26000 & 1.3 & 10.0 & 10.0 & \\
   & PL & 26000 & 1.3 & --- & 11.8 & \\
&&&&&& \\
\multicolumn{2}{l}{\bf BULGE} &&&&& \\
&&&&&& \\
Model & Type & $N_B$ & $M_B$ & $b_B$ & $R_{B}$ & \\
{\bf m1} & PL & 2000 & 0.1 & 0.375 & 10.0 & \\
{\bf m2} & PL & 2000 & 0.1 & 0.375 & 10.0 & \\
{\bf m3} & PL & 2000 & 0.1 & 0.375 & 10.0 & \\
{\bf s1} & PL & 2000 & 0.0997 & 0.375 & 10.0 & \\
{\bf mb} & --- & --- & --- & --- & --- & \\
{\bf mh} & --- & --- & --- & --- & --- & \\
&&&&&& \\
\multicolumn{3}{l}{\bf COMPANION} &&&& \\
&&&&&& \\
Model & Type & $N_C$ & $M_C$ & $b_C$ & $R_{C}$ & \\
{\bf cs} & PL & 4000 & 0.1987 & 0.195 & 3.0 & \\
{\bf csd} & PL & 8000 & 0.3974 & 0.195 & 3.0 & \\
{\bf csh} & PL & 2000 & 0.09935 & 0.195 & 3.0 & \\
{\bf c1} & PL & 800 & 0.04 & 0.3333 & 3.0 & \\
\end{tabular}
\end{table}

The setting up procedure follows approximately
that of Barnes (1988). We 
first construct the halo and bulge components separately and then superpose 
them and allow them to relax together. Then we choose positions for the disc 
particles, using the prescribed density profiles, and tabulate the forces 
from this distribution of
particles on a grid. This field is slowly imposed 
on the spheroid particles, in 
addition to their own field, to allow the halo-bulge system to relax in the 
total field. After the disc field has reached its 
final amplitude, we continue the evolution until the spheroid reaches
equilibrium. The remaining step
is to give the disc particles their initial velocities. For this we 
use the chosen velocity 
dispersions, take 
into account asymmetric drift corrections, and use the information from 
the relaxed halo-bulge component for the calculation of the initial tangential 
velocity of the disc particles. At this point we are ready to start the 
si\-mu\-lation by superposing the spheroid and disc distributions and adding
the companion.

Model {\bf mb} has no bulge and is bar unstable. 
The setting up procedure was the same as described above, except that there 
was no bulge component. In this case, however, before introducing the companion,
we let the galaxy 
evolve, first enforcing axisymmetry, and then allowing full freedom of all the 
particles, until the bar forms, and only then did we introduce the 
companion. In other words, in this case the companion will be perturbing a
barred galaxy. 
Thus model {\bf mb} is described in 
Table~\ref{tab:models} by two 
lines. The first one gives the parameters describing the individual 
components before they interacted, as for the previously described models. 
The second line gives information on the 
target galaxy at the time the companion is introduced. As $b_D$ 
($z_0$) we give the radius (height) containing 
the same percentage of the disc mass as $b_D$ ($z_0$) for the unperturbed
disc. No equivalent quantity can be given for $b_H$ since,
because of the adopted truncation, $b_H$ contains 100\% of the Plummer sphere 
mass. As $R_{D}$ and $R_{H}$ we give 
the radii containing 99\% of the disc and halo mass, respectively. 

To obtain model {\bf mh} we took the barred galaxy described above and 
redistributed the disc particles randomly in azimuth. For this reason both 
lines describing this model in Table~\ref{tab:models} 
contain identical information (except of course for the 
descriptions ``hot" and ``barred") to those describing model {\bf mb}. This model 
has high values of the radial velocity dispersion in the disc (a density
weighted mean $Q$ of the order of 1.8) and thus allows 
us to perturb a hot disc in a bulgeless galaxy. We have also evolved this model 
without a companion for a $\Delta t=120$, which is considerably longer than 
the time 
we follow the target disc after the perturber has been put in, and find that 
no noticeable 
bar component formed, although Fourier analysis of the particle positions 
shows a weak signal at later times 
that could develop into a bar if the run had been evolved further. 
However this signal is negligible during approximately 65 units of time 
which allows us to consider simulations with this target as perturbations 
of a nonbarred disc galaxy (in the two simulations where this model
was used, the companion hits the disc at $t=26$ and $t=8$ units of time).

The companion is also modelled as a Plummer sphere and the parameters 
describing it, namely the number of particles $N_C$,
its mass $M_C$, its characteristic radius $b_C$ and its 
cutoff radius $R_{C}$, are listed in the last section of
Table~\ref{tab:models}. For brevity we will often refer to companion 
{\bf cs} as standard, while {\bf csd} and {\bf csh} will be referred 
to as having double or half the standard mass respectively.

Runs are labelled S1 to S4, R1 to R12 and C55 to C99 and are listed in Table 
\ref{tab:components}. In runs 
R1 to R12 and C55 to C99 all particles in the target galaxy have the same 
mass, so that the ratio of 
number of particles equals the mass ratio between the different components. 
In runs S1 to S4 the masses of disc, halo, and bulge particles are
$4\times 10^{-5}$, $3.57\times 10^{-5}$ and $5\times 10^{-5}$
respectively. Thus the mass ratio of the 
components is \hbox{$D:H:B = 1:3.34:0.31$}. The mass of the particles 
in  companions {\bf cs}, {\bf csh} and {\bf csd} is $4.9657\times 10^{-5}$ and that of 
the particles in companion {\bf c1} is $5\times 10^{-5}$.

\begin{table}
\centering
\caption{Components, type of the impact and code used in the runs }
\label{tab:components}
\begin{tabular}{@{}ccccc@{}}
RUN & MODEL & COMPANION & IMPACT & CODE \\
S1 & {\bf s1} & {\bf cs} & PS & treecode \\
S2 & {\bf s1} & {\bf cs} & PF & treecode \\
S3 & {\bf s1} & {\bf cs} & CS & treecode \\
S4 & {\bf s1} & {\bf cs} & CF & treecode \\
R1 & {\bf m1} & {\bf cs} & CS & treecode \\
R2 & {\bf m1} & {\bf cs} & CF & treecode \\
R3 & {\bf m2} & {\bf cs} & CS & treecode \\
R4 & {\bf m3} & {\bf cs} & CS & treecode \\ 
R5 & {\bf mh} & {\bf cs} & CS & treecode \\
R6 & {\bf mh} & {\bf cs} & CF & treecode \\
R7 & {\bf mb} & {\bf cs} & CS & treecode \\
R8 & {\bf mb} & {\bf cs} & CSC & treecode \\
R9 & {\bf mb} & {\bf cs} & PSB & treecode \\
R10 & {\bf m1} & {\bf c1} & CF & treecode \\
R11 & {\bf m1} & {\bf c1} & C01 & treecode \\
R12 & {\bf m1} & {\bf c1} & C02 & treecode \\
C55 & {\bf m1} & {\bf cs} & CS & grape \\
C57 & {\bf m1} & {\bf csd} & CS & grape \\
C58 & {\bf m1} & {\bf csh} & CS & grape \\
C59 & {\bf m1} & {\bf cs} & CVS & grape \\
C61 & {\bf m1} & {\bf csd} & CVS & grape \\
C63 & {\bf m1}(Q=1.1) & {\bf csd} & CS & grape \\
C64 & {\bf m1}(Q=1.2) & {\bf csd} & CS & grape \\
C85 & {\bf m2} & {\bf cs} & CS & grape \\
C86 & {\bf m2} & {\bf csd} & CS & grape \\
C88 & {\bf m3} & {\bf cs} & CS & grape \\
C89 & {\bf m3} & {\bf csd} & CS & grape \\
C91 & {\bf m1}(Q=1.5) & {\bf csd} & CS & grape \\
C99 & {\bf m1} & {\bf cs} & CS & grape \\
\end{tabular}
\end{table}

The units of length and time are 3 kpc and $10^7$ years respectively, and 
$G=1$. Using this normalisation the units of mass, velocity and volume density 
are \hbox{6 $\times$ 10$^{10}$ M$_{\odot}$},
\hbox{293 km/sec}, and
\hbox{2.22 M$_{\odot}$/pc$^3$}
respectively.

\subsection{Simulations }
\indent

Table~\ref{tab:components} lists all the runs discussed in this paper 
(co\-lumn~1), together with the model
used for the target and the companion (columns~2 
and 3) and, in
column~4, an indication of the initial conditions used for the
companion's orbit. In the last column we give the code used in
the force calculation.
The corresponding initial positions (columns~2 to 4) and
velocities (columns~5 to 7) of the companion are given in computer units in
Table~\ref{tab:initial}. The last two columns of this table give the ratio of
the amplitude of the initial velocity to the escape velocity, calculated
by assuming all the mass in the target galaxy is at a point in its center.

The initial positions and velocities 
have been obtained as follows: we first decide the impact point and
velocity for model {\bf s1} and companion {\bf cs}, then calculate the orbit
backwards in time for a time interval sufficiently long that the distance 
between the center of the target galaxy and the companion is longer 
than $1.4 R_{H}$, while keeping the
target galaxy particles fixed. The final positions and velocities are then
used as the initial positions and velocities of the companion for initial
conditions CS, CF, PS and PF, corresponding to simulations S1 to S4, R1 to 
R7, C55 to C58, C63 to C99.
Of course when the full blown simulation is run, the companion
never hits the disc exactly at the impact point and with the impact velocities
initially chosen, partly due to the evolution of the disc and partly
because of the diffe\-rences between the discs {\bf m1}, {\bf m2}, {\bf m3},
{\bf mh} and {\bf mb} on the one hand and disc {\bf s1} on the other.
Nevertheless, Table~\ref{tab:impact} shows that the differences are not 
large 
and that CF initial conditions correspond roughly to central and fast
encounters, CS to central and slow ones, PF to peripheral and fast, and PS to
peripheral and slow ones. The initial conditions for run R8 (CSC) are similar to
those of run R7, except for a slight spatial shift to make the impact position 
of the
companion as near as possible to the density center of the disc. For run R9
the procedure is the same, except that we have used disc {\bf mb} when
calculating the orbit backwards to get a slow impact on the bar semimajor 
axis. The initial conditions C01 and C02 
are such that the companion starts at rest from a point 12 computer units 
from the center of the target galaxy, either on the $z$ axis (C01), or
$30^\circ$ from it (C02). Finally CVS is a central and vertical passage
whose initial radius and velocity amplitude are the same as those of CS.

\begin{table*}
\centering
\begin{minipage}{140mm}
\caption{Initial conditions}
\label{tab:initial}
\begin{tabular}{@{}ccccccccc@{}}
& \multicolumn{3}{c|}{POSITION} & \multicolumn{3}{c|}{VELOCITY} & 
\multicolumn{2}{c|}{$\vert V/V_{esc}\vert$} \\
IMPACT & $x$ & $y$ & $z$ & $V_x$ & $V_y$ & $V_z$ & S runs & R or C runs \\
PF & -10.02 & 10.04 & 20.30 & 1.04 & -1.05 & -2.31 & 7.88 & \\
PS & 0.0 & 10.0 & 10.0 & -0.04 & 0.0 & 0.0 & 0.08 & \\
CF & -8.98 & 9.39 & 18.08 & 1.03 & -1.08 & -2.07 & 6.96 & 5.67 \\
CS & -10.01 & 19.07 & 17.39 & 0.33 & -0.63 & -0.56 & 2.75 & 2.35 \\
CSC & -10.11 & 19.17 & 17.39 & 0.33 & -0.63 & -0.56 & & 2.38 \\
CVS & 0.0 & 0.0 & 27.68 & 0.0 & 0.0 & -0.90 & & 2.36 \\
PSB & -4.74 & 17.25 & 15.19 & 0.10 & -0.59 & -0.51 & & 1.91 \\
C01 & 0.0 & 0.0 & 12.0 & 0.0 & 0.0 & 0.0 & & 0.0 \\
C02 & 0.0 & 6.0 & 10.39 & 0.0 & 0.0 & 0.0 & & 0.0 \\
\end{tabular}
\end{minipage}
\end{table*}

\begin{table*}
\centering
\begin{minipage}{140mm}
\caption{Values at the impact}
\label{tab:impact}
\begin{tabular}{@{}ccccccccc@{}}
RUN & 
RING & 
$T_i$ & 
$\theta_i$ & 
$R_i$ &
$V_{C_{R_{i}}}$ &
$V_{C_{z_{i}}}$ &
$\vert V_{C_{R_{i}}}$/$\sigma_{R_{i}}\vert$ &
$\vert V_{C_{z_{i}}}$/$\sigma_{z_{i}}\vert$ \\
S1 & y & 42 & 45 & 0.50 & -0.5 & -0.8 & 2.3 & 4.5 \\
S2 & y & 8 & 56 & 1.20 & -1.6 & -2.4 & 11.6 & 21.8 \\
S3 & y & 26 & 43 & 0.04 & 1.1 & -1.2 & 3.8 & 4.6 \\
S4 & y & 9  & 54 & 0.03 & 1.6  & -2.5 & 5.8 & 9.2 \\
R1  & y & 26 & 43 & 0.07 & 1.2 & -1.1 & 5.9 & 4.4 \\
R2  & y & 8 & 54 & 0.05 & 1.6 & -2.3 & 10.7 & 9.6 \\
R3 & y & 26 & 42 & 0.05 & -1.1 & -1.1 & 10.0 & 7.2 \\
R4 & y & 27 & 42 & 0.17 & -1.0 & -1.1 & 3.3 & 5.4 \\
R5  & y & 26 & 42 & 0.07 &  1.1 & -1.1 & 3.6 & 4.6 \\
R6  & y & 8 & 55 & 0.07 & 1.5 & -2.4 & 4.5 & 8.5 \\
R7 & y & 26 & 45 & 0.21 & -1.1 & -1.0 & 3.3 & 4.5 \\
R8 & y & 26 & 43 & 0.24 & 0.3 & -1.2 & 0.9 & 4.7 \\
R9 & y & 25 & 42 & 0.91 & 1.0 & -1.3 & 3.5 & 6.5 \\
R10 & n & 4 & 89 & 0.02 & 0.0 & -3.2 & 0.0 & 13.2 \\
R11 & y? & 34 & 88 & 0.03 & 0.0 & -1.2 & 0.0 & 5.5 \\
R12 & y? & 34 & 58 & 0.06 & -0.6 & -1.2 & 2.8 & 4.5 \\
C55 & y & 26 & 42 & 0.04 & 1.1 & -1.1 & 5.9 & 4.4 \\
C57 & y & 26 & 42 & 0.06 & 1.1 & -1.3 & 3.7 & 5.3 \\
C58 & y & 26 & 41 & 0.08 & 1.1 & -1.0 & 7.2 & 4.1 \\
C59 & y & 26 & 89 & 0.02 & 0.0 & -1.6 & 0.0 & 6.9 \\
C61 & y & 26 & 89 & 0.01 & 0.0 & -1.8 & 0.0 & 7.3 \\
C63 & y & 26 & 41 & 0.07 & 1.1 & -1.3 & 4.6 & 5.1 \\
C64 & y & 26 & 42 & 0.07 & 1.1 & -1.3 & 4.5 & 5.1 \\
C85 & y & 26 & 41 & 0.01 & 1.1 & -1.0 & 6.3 & 4.6 \\
C86 & y & 26 & 41 & 0.08 & 1.1 & -1.1 & 4.7 & 5.5 \\
C88 & y & 26 & 42 & 0.09 & 1.0 & -0.9 & 5.2 & 6.2 \\
C89 & y & 26 & 42 & 0.07 & 1.0 & -1.0 & 5.9 & 5.3 \\
C91 & y & 26 & 45 & 0.06 & 1.0 & -1.2 & 4.0 & 7.3 \\
C99 & y & 26 & 41 & 0.08 & 1.1 & -1.1 & 5.8 & 4.6 \\
\end{tabular}
\end{minipage}
\end{table*}

Thus Table~\ref{tab:components} gives at a glance rough information about the run description. For example the line describing run C55 tells us that the target had a standard  disc (second column), the companion was standard (third column) and the impact was central and slow (fourth column). Similarly for run C86 we see that the target has the more extended disc {\bf m2}, a double mass companion and a central and slow impact.

In Table~\ref{tab:impact} the first column gives the label of the run and
the second column indicates whether a ring was formed (y) or not (n). 
Runs with rings which are not
clearly outlined are denoted with a y?. Columns~3 to 9 give the time
of impact measured from the time the companion was introduced, 
the impact angle 
(i.e., the angle between the plane of the disc and
the orbit of the companion at impact), the distance of the impact point from
the center of mass of the target galaxy, the radial and vertical 
velocity 
component of the companion at impact, and the ratios of the absolute values
of these components to the velocity dispersion along the same direction
and measured at the impact position. The angle is measured in degrees 
and the other quantities in computer units. As impact time we list 
the time step immediately preceding or 
following the impact, for which the companion was nearest to the $z=0$ plane 
and for which the information on the particle position and 
velocities exists. Since all particle positions and velocities were stored every 
two computer time units for runs C55 to C99 and every time unit for the 
remaining runs (see next section), the precision for the impact time is $\pm$ 1
for runs C55 to C99 and $\pm$ 0.5 for all other runs. The intersection between the 
$z=0$ plane and the line connecting the positions of the center of mass of 
the companion at the time steps saved immediately before and after impact
defines the impact angle and the distance of the impact point from the 
center. The velocities of the last four columns are measured 
by interpolation at the time 
when the companion crosses the disc.

\subsection{Numerical codes }
\indent

For all runs starting with an R or an S
the evolution has been
followed using a vectorised version (Hernquist 1990) of the
treecode (Barnes \& Hut 1986). For the R models we
use a tolerance parameter of $\theta = 0.7$, and for the S runs
$\theta=1$. In both cases we include quadrupole terms in the force
calculations to increase the accuracy.
The softening parameter has been taken to
be the same for all particles and equal to 0.066666 for 
the R runs and 0.05 for the S
runs, and the time step 
equal to 0.025 and 0.05 respectively, both measured 
in computer units. 
For the parameters of the R runs and 44000 particles one time step
took roughly 15 seconds, the precise value depending on the configuration.
These parameters also
ensure an adequate energy conservation. If we 
measure the energy conservation using the minimum and maximum values obtained 
during the runs, we find an accuracy of order 0.1\%, the highest 
deviations
being often found at times when the companion crosses the disc and 
becomes very concentrated. If we consider only the initial and final values of 
the energy, then we obtain an energy conservation better than 0.05\%.

All these simulations have been run for 80 computer units, or equivalently 
$8\times
10^8$ years, from the time the companion is introduced. This time span is
sufficient, since rings formed by infalling small companions prove to be
short lived structures. The positions and velocities of all particles in the
simulation have been 
saved every integer value of the time measured in computer
units. 

\begin{figure*}
  \vspace{13.00cm}
\includegraphics{fig01.ps}
  \caption{Snapshots from simulation C59 showing the formation and evolution
    	of the first and second ring.
	We plot the x, y positions of 
	all disc particles in computer units. Times are marked in the 
	upper left corner of each frame and are 
	measured from the time of impact, in computer units, i.e., with the
        adopted normalisation, in units of $10^7$ years.}
  \label{fig:c59xy}
\end{figure*}

After this project had been started, our group acquired a GRAPE system 
(Okumura et al. 1992, Ebisuzaki et al. 1993, Okumura et al. 1993) 
consisting of 5 GRAPE-3AF boards coupled via an Sbus/VMEbus converter to a 
SPARC station 10/512. A detailed description of this system and of its 
performances will be given elsewhere. Here we will only mention that, for
the direct summation code used in this paper,
properly parallelised over the 40 chips,
it gives a sustained speed slightly higher than 20 
Gflops, so that one time step for 44000 particles takes only 3 seconds.
With this system we have calculated runs C55 to C99, using direct 
summation. The time
step have been chosen equal to 0.015625 and the softening 0.0625, both
in computer units. Again we find that the highest errors 
in the energy correspond
to the time when the companion crosses the disc. Nevertheless the
energy conservation, calculated from the initial and final values,
is better than or of the order of 0.2 \%. 

All simulations made with GRAPE have been run for 100 computer time units, or
equivalently $10^9$ years, from the time the companion is
introduced. The positions and velocities of all particles in
the simulations have been saved every 2 computer time units, or
equivalently $2 \times 10^7$ years.

	The advantages  of these two approaches are multiple. 
All components, including the halo and the companion,
are described fully self-consistently. They allow a good 
resolution for all geometries and configurations. No extra forces need be 
introduced to account for the change of reference frame due to the 
companion,
as in the case for grid codes.
Finally the orbit of the companion comes naturally out of the simulation and 
does not have to be assumed or pre-calculated.

Since we include in this paper simulations made with different codes
and computers --- treecode on CRAY and direct summation on GRAPE --- we
have run a few cases in both ways in order to make comparisons. Thus
run R1 has identical initial conditions with run C55. C99 is also 
identical to C55 except that the initial positions of the disc particles
have been randomised in azimuthal angle. We compared the radius of the
ring as  a function of angle and time and found very good agreement
between the three cases. Furthermore the differences between C55 and C99
are of the same order as those between R1 and C55, or R1 and C99.

\section{A nonbarred target galaxy }
\label{sec:nonbar}

\subsection{The fiducial case: A central vertical impact }
\indent

The rings in our simulations show very different morphologies, depending on 
the mass and trajectory of the companion but also on the target disc. A good 
point to start the descriptions and comparisons is run C59, which 
is a central 
vertical impact of the standard companion ({\bf cs}) on the standard disc 
({\bf m1}), and
whose most relevant part is shown in Fig.~\ref{fig:c59xy}.
The first ring starts right after impact and is, all through its evolution, 
rather symmetric and near-circular. It expands initially very rapidly but the 
expansion rate slows down somewhat with time. The second ring is less 
circular and also expands less fast.

	One can get a clear impression of how the rings evolve with time by
using an $r = r (t)$ plot, as given in Fig.~\ref{fig:r=rtb}, where $r$ is the
radius of a particle in the disc component, measured from the center of mass
of the target galaxy and $t$ the time measured from the beginning of the 
simulation. Particles in the halo and bulge are not plotted, and,
for cla\-ri\-ty, we only display 2000 of the disc particles at each time. 
Strictly speaking
it is not possible to get $r=r(t)$ from numerical simulations since these
give information about the positions of the particles only at discrete times.
We have thus used the following artifact, quite similar to what is used for
greyscale plots. We saved the values of the positions of all particles in the
simulation very frequently, namely every unit of time 
for S and R runs and 2 units of time for C ones (in computer units).
Plotting this information on the $(r,t)$ plane would give infinitely thin 
strips of points, which
would not allow us to see any evolution. We have thus smeared out the
information by placing every particle at its radius and at a time chosen at
random between $t$ and $t+1$ or $t$ and $t+2$. 
For treecode runs where the particle 
coordinates are saved every unit of time and where the rings happen to 
have re\-la\-tively low expansion velocities, this procedure gives a very smooth 
figure, allowing us to follow closely the evolution of the ring (e.g. 
Fig.~\ref{fig:r=rtc}).
For simulations with large ring expansion velocities and for which the 
particle coordinates are saved every two time units, this procedure gives a 
more step-wise appearance (e.g. lower panel of Fig.~\ref{fig:r=rtd}), yet 
clear enough to allow us to draw conclusions.
The upper panel of Fig.~\ref{fig:r=rtb} shows the data for 
our fiducial simulation. One can clearly see the formation of the two first rings.
 One
can also follow the expansion of the rings and see that it is faster for the
first than for the second ring, and also that it slows down with time.

Insight into the ring formation and evolution can be obtained by putting in 
equations figure 4 of Lynds \& Toomre (1976) or figure 6 of Toomre (1978). 
This is easily done by following the motion of particles initially in circular 
orbits and perturbed by the infalling companion. Binney and Tremaine (1987) 
model the target galaxy and the companion as two identical isothermal spheres,
 use the impulse approximation to calculate the effect of the companion, which 
is assumed to have a constant velocity, and neglect collective effects in the ring\footnote{In a similar way, but taking into account that the companion is 
a Plummer sphere, we obtain for the radius at time $t$ of a star initially at 
radius $R_o$

${\displaystyle R(R_o,t)=R_o-\frac{2GM_c}{V\kappa} \frac{R_o}{R_o^2+b_c^2} sin(\kappa t)}$

\noindent
where $V$ is the velocity of the companion, assumed constant, and where the 
epicyclic frequency is measured at the radius $R_o$. This equation shows that 
slower passages, or more massive companions create larger displacements of the 
orbits. This predicts higher density in the rings for such passages, and, as we 
will see in the next sections, this is indeed borne out by our simulations.}.
Using this equation to calculate the perturbed surface density we see that the 
density enhancement moves outwards with a velocity which is constant if the 
rotation curve is constant with radius. On the other hand if the rotation 
curve decreases with radius, as in the examples of Lynds \& Toomre (1976) 
and Toomre (1978) - where the drop was keplerian -, this velocity decreases 
with radius. Such a decrease can also be seen in our figures \ref{fig:r=rtb}, \ref{fig:r=rtc}, and \ref{fig:r=rtd}. The 
reason in this case is not the form of the rotation curve, which, within 
the disc, is nearly constant, but collective effects, which have been 
neglected in the above analysis. This can be seen by calculating numerically 
the group velocity (cf. Toomre 1969) from the $m=0$ Lin-Shu-Kalnajs dispersion 
relation (Kalnajs 1965, Lin \& Shu 1966). 

\subsection{Non-vertical passages }
\indent

\begin{figure}
  \vspace{10.00cm}
\includegraphics{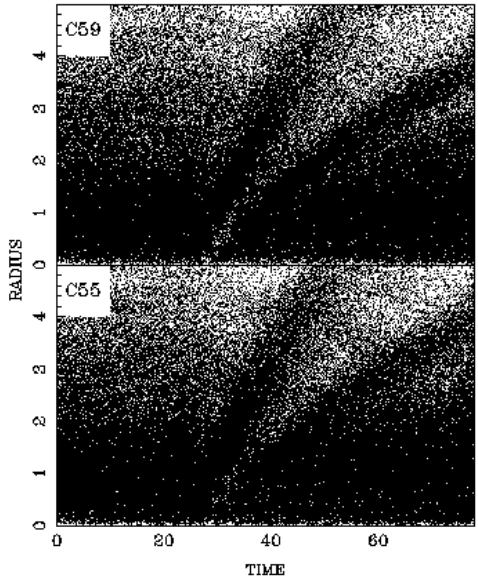}
  \caption{Comparison of the $r=r(t)$ plots, as described in the text, 
for a vertical (upper panel) and an oblique (lower panel)
impact. The label of the simulation is given in the upper 
left corner of each panel. Both radii and times are in computer
units.}
  \label{fig:r=rtb}
\end{figure}

Run C55 has the same target and companion as run C59
but the impact is not vertical but 
rather at an angle of roughly $45^{\circ}$. Both the first and 
the second (although 
the latter to a lesser extent) rings are more eccentric and broader than 
those generated 
by the vertical impact and their axial ratio does not vary much with time.
The major axis of the ring rotates.
Fig.~\ref{fig:r=rtb}
compares the $r=r(t)$ plots of these two simulations  
and shows that the differences are small. 

\subsection{Fast and slow passages }
\indent

\begin{figure}
  \vspace{10.00cm}
\includegraphics{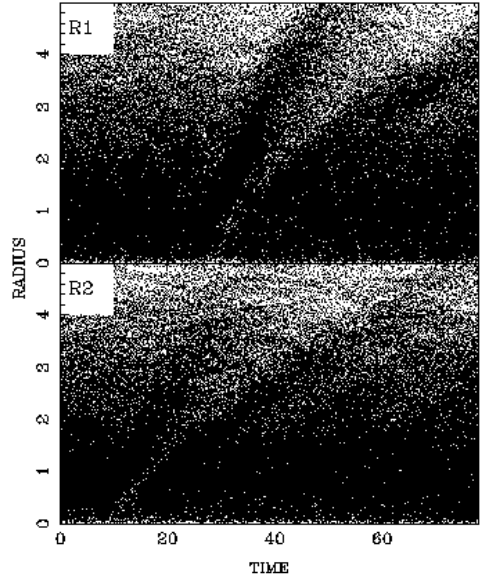}
  \caption{Comparison of the $r=r(t)$ plots
for a slow (upper panel) and a fast passage (lower panel).
The label of the simulation is given in the upper 
left corner of each panel. Both radii and times are in computer
units.}
  \label{fig:r=rtc}
\end{figure}

The evolutions of simulations with slow (R1) and fast (R2) passages do not differ much. Nevertheless the $r=r(t)$ plot (Fig.~\ref{fig:r=rtc}) shows that for the fast passage the rings are less intense, do not reach the edge of the disc and 
expand less fast. Also in the case of the faster passage there is less time 
during which the two rings coexist, because the first ring fades away faster.

\subsection{Varying the mass of the companion }
\indent

\begin{figure*}
  \vspace{17.00cm}
\includegraphics{fig04.ps}
  \caption{Comparison of three simulations with different perturber masses. 
The left panels correspond to simulation C58, where the companion has half 
the standard mass, the middle ones to simulation C55, where the companion has 
the standard mass, and the right ones to C57 with a double mass perturber. 
The arrows plotted in the right column show the direction of the 
companion's velocity at impact and start off from the impact point.
Time increases from top to bottom and is marked in the top left corners of 
the left panels, all panels in the same row corresponding to the same time.}
  \label{fig:585557xy}
\end{figure*}

The effect of changing the perturber mass is quite sizeable, as can be seen 
from Fig.~\ref{fig:585557xy}, where we compare runs with identical target 
discs but companions of half the standard mass (C58), standard mass (C55) 
and double the standard mass (C57).

\begin{figure}
  \vspace{10.00cm}
\includegraphics{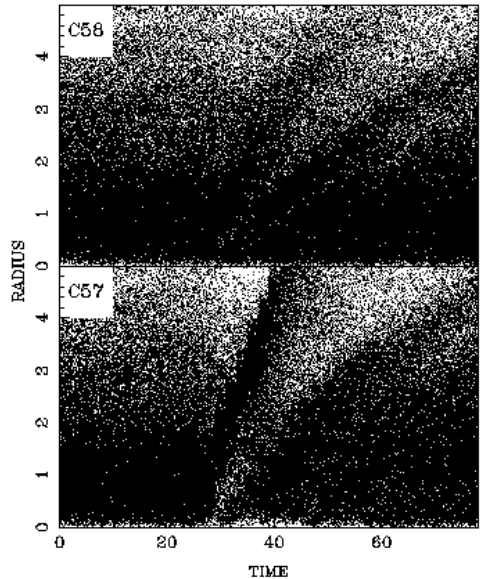}
  \caption{ Comparison of $r=r(t)$ plots for a companion of low mass
(upper panel) and of high mass (lower panel).
The label of the simulation is given in the upper 
left corner of each panel. Both radii and times are in computer
units.}
  \label{fig:r=rtd}
\end{figure}

The companion-to-disc mass ratios in these cases are 0.25, 0.5 and 1.  
respectively, and the companion-to-target mass 0.05, 0.1 and 0.2 respectively . 
Both the expansion velocity of the first ring 
and its intensity increase considerably with the mass of the companion. This 
is clearly seen from the three upper rows, which correspond to early times in 
the evolution. The last row corresponds to a much later time. The first ring 
has faded out in the low mass encounter, while being still present in the 
other two, and is particularly clear in the high mass case. At this time the 
second ring is still clear in all three simulations, although its structure is 
rather different. Another important difference is the existence of spokes in 
the high mass case, which will be discussed in section \ref{sec:spokes}, and the
sizeable expansion of the disc extent, which can be seen by comparing the top 
and bottom rows for the three simulations. Rings are more symmetric and 
nearer to circular in low mass encounters than in higher mass ones. There is 
also a general trend that more massive companions create wider, more intense 
rings. Similar results can be seen by comparing simulations C59 and C61, not 
shown here.

\begin{figure*}
  \vspace{17.00cm}
\includegraphics{fig06.ps}
  \caption{Comparison of three simulations with the same standard mass 
companion but 
different target discs, described as 
\mxum (C55), \mxdois (C85) and \mxtres (C88) in the 
text. Description of the layout as for Fig.~2.}
  \label{fig:558588xy}
\end{figure*}

\begin{figure*}
  \vspace{17.00cm}
\includegraphics{fig07.ps}
  \caption{As for Fig.~6 but for a 
     companion of double mass.}
  \label{fig:578689xy}
\end{figure*}

\begin{figure}
  \vspace{18.00cm}
\includegraphics{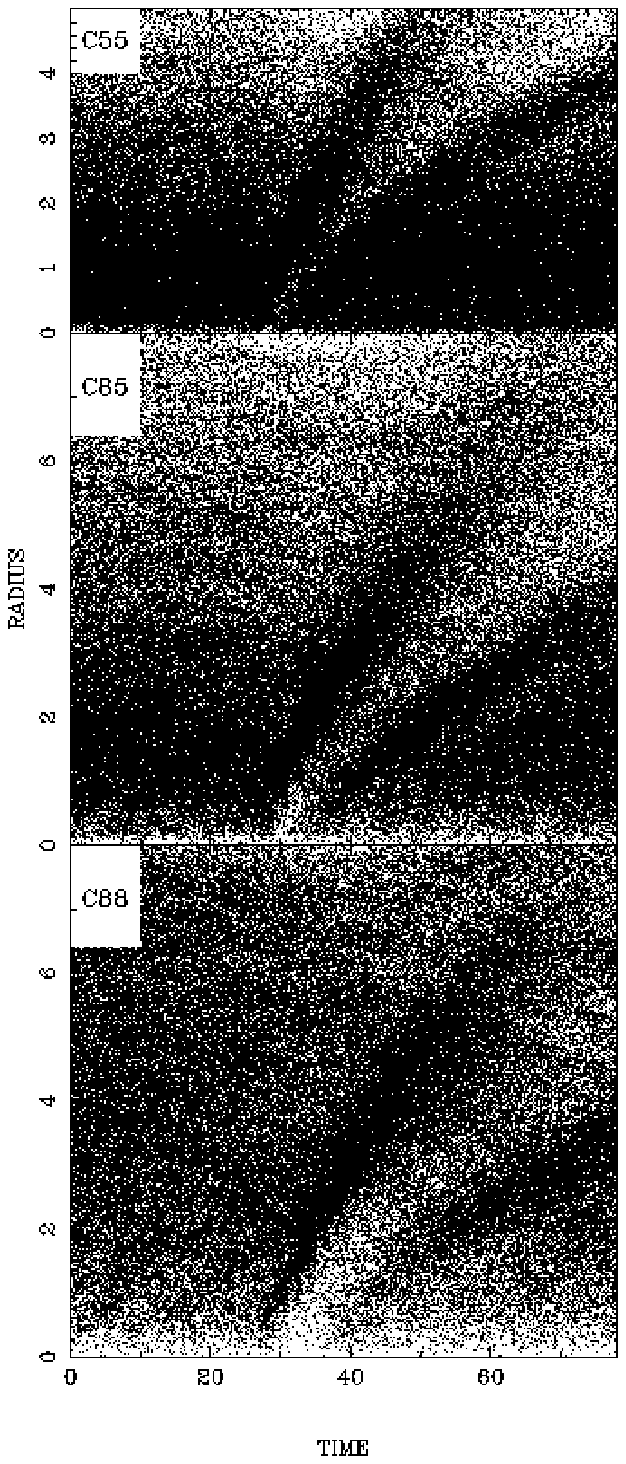}
  \caption{Comparison of $r=r(t)$ plots for three simulations with target discs of 
different scalelengths, the smallest scalelength being at the top and the 
biggest one at the bottom.}
  \label{fig:r=rt2}
\end{figure}

Fig.~\ref{fig:r=rtd} compares the $r=r(t)$ plots for the impact of the low mass 
companion (C58) to that of the high mass one (C57) and shows that there
is an important increase of the expansion velocity and of the ring 
intensity with the companion mass. 

If the companion has a very low mass, then rings either do not form, or are too weak to be clearly seen. This is the case for simulations R10, R11 and R12 for which the 
companion has a mass of only 0.04 computer units, or $M_C/M_D=0.1$ and 
$M_C/M_G=0.02$, where $M_C/M_D$ and $M_C/M_G$ are the ratios of the mass in 
the companion to that in the disc and target galaxy respectively.

\subsection{Different target discs }
\indent

Striking differences can be seen when one varies the scalelength of the 
target disc, i.e., when one considers in turn discs 
{\bf m1}, {\bf m2} and {\bf m3}. 
Our simulations allow us to construct three such sequences, two with the 
standard companion mass, of which one is shown in Fig.~\ref{fig:558588xy},  
and the other with double mass, shown in Fig.~\ref{fig:578689xy}. As 
expected from swing amplifier theory (Toomre 1981), the more extended discs 
show before the impact a spiral structure with higher arm multiplicity. 
Then, when the ring forms, it is much more homogeneous in the 
standard disc than in the extended ones, where it is very irregular and 
patchy and, in the most extended disc case, ends up looking like a polygon 
rather than an oval (cf.~the third row, third column panels of 
Figs.~\ref{fig:558588xy} and \ref{fig:578689xy}). At yet later times the 
first ring breaks up in many spiral arcs and segments, as can be seen in the bottom right panel of Fig.~\ref{fig:558588xy}.

Fig.~\ref{fig:r=rt2} compares the $r=r(t)$ results for the three 
si\-mu\-lations shown in Fig.~\ref{fig:558588xy}. It gives the 
impression that the more
intense rings are formed in the more compact disc, an impression
which will be confirmed by a measurement in section \ref{sec:ampl}.

\subsection{Asymmetries }
\indent

Since the impact in most of our simulations is not vertical we expect some 
asymmetries in the resulting rings. Indeed we found that the first rings 
obtained from double mass companions can look quite asymmetric, 
particularly towards the end of their lifetime, but not the rings formed by 
standard or half mass companions. This can be seen for example in the last 
row of Fig.~\ref{fig:585557xy}, where it is clear that the asymmetry 
increases with companion mass although the initial conditions for the three 
simulations are the same. It can also be seen by comparing 
Figs.~\ref{fig:558588xy} and \ref{fig:578689xy}. 
The arrows plotted in the right hand column of Fig.~\ref{fig:585557xy} show 
the direction of the companion's velocity at impact and
start off from the impact point. We note that the highest
density point stays near the impact position and the direction of
least expansion is not far from the direction to which the arrow is 
pointing. In the last plotted time the direction of the impact 
roughly coincides with the major axis of the inner ring and inner oval.
These structures, however, rotate, so this will not hold for other 
moments.

Asymmetries can also be obtained of course with peripheral impacts. 
An example is given in Fig.~\ref{fig:s1xy}, which
shows four instants from the evolution of simulation S1. The ring that forms is 
offcentered and non-circular, and the position of the ``nucleus" with respect 
to the ring changes with time. There is also one single spoke 
or spiral 
feature, which is diffuse and lasts until about $t=14$. At roughly $t=15$ the 
companion crosses the disc for a second time and it is this passage that 
might be responsible for the demise of the spoke. In the later stages of 
the evolution the spoke, then the nucleus, and finally the ring itself, become 
quite deformed. 

For our second simulation with a peripheral impact (S2) the impact is too 
peripheral and/or too fast 
(cf.~Table~\ref{tab:impact}) to produce 
a ring as clear as those of the previous case.

\subsection{Spokes }
\label{sec:spokes}
\indent

The Cartwheel galaxy (A0035--335) is one of the best studied examples of a 
ring galaxy. Detailed high contrast photographs (e.g., Plate 1, Davies
\& Morton 1982; or Fig. 4, Toomre 1978)
show, in addition to the inner and outer ring, a set of short radial 
arcs, or segments,
frequently referred to as spokes. These features have been reproduced
in a transient manner in some of our simulations
and we describe them in this section. They are located between the 
inner and outer ring. As the outer ring expands material which at 
some point was in it falls back towards the center of the galaxy. 
Clumps and inhomogeneities in the ring are also present in this 
material and trigger swing amplification (Toomre 1981). Thus spokes 
can be understood in the same way as the sections of spiral arms 
seen in flocculent galaxies.

\begin{figure*}
  \vspace{12.00cm}
\includegraphics{fig09.ps}
  \caption{Four characteristic times of the evolution of 
         run S1. Note the spoke around $t=11$.}
  \label{fig:s1xy}
\end{figure*}

\begin{figure}
  \vspace{7.00cm}
\includegraphics{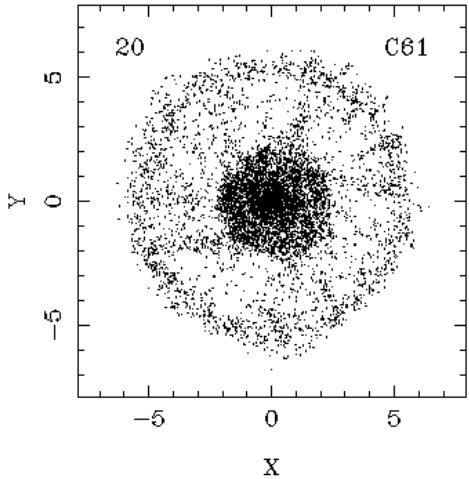}
  \caption{One example of spokes from simulation C61.}
  \label{fig:c61}
\end{figure}

An example of such spokes in our simulations are seen clearly in  
Fig.~\ref{fig:c61}, a frame from the vertical central slow
passage in run C61. They appear 
between 
the outer and inner ring, the presence of both being necessary. They
are trailing, nearly straight and last for a couple of $10^8$ years.
Run C57, which
has the same target and companion, but an impact at $42^{\circ}$, also
shows spokes, one quite massive and the others not so well defined.

A problem that needs to be further 
considered is what the mass ratio between the companion and the target 
should be for such features to appear. In our simulations encounters 
with a companion of mass twice 
the standard, i.e., a mass equal 
to that of the disc or 20\% of the mass of the galaxy, produce spokes at 
some stage of their evolution, whereas 
this is sometimes but not always the case for encounters with standard 
mass companions, as 
can be seen from Figs.~\ref{fig:c59xy}, \ref{fig:585557xy}, 
\ref{fig:558588xy}, \ref{fig:578689xy} and \ref{fig:s1xy}. Seen the 
observational uncertainties in the
calculation of the mass of the Cartwheel and its companions (Appleton \& 
Struck-Marcell 1996), these numbers tend to suggest that one of the companions, 
and in particular G2 (following the notation of Higdon 1996), could be 
responsible for the structures in the Cartwheel (cf. also Struck-Marcell 
\& Higdon 1993). A confirmation, however, would necessitate more elaborate 
modelling and, in particular, a target rotation curve resembling that of 
the Cartwheel.

\begin{figure}
  \vspace{6.00cm}
\includegraphics{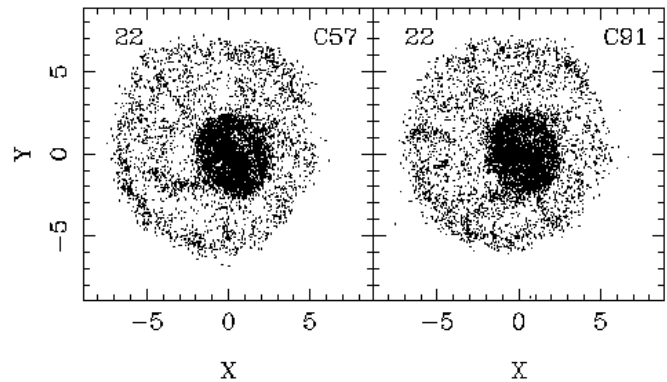}
  \caption{Comparison of two simulations with identical initial conditions, 
          except for the initial $Q$ value, which is 1 for run C57 
          and 1.5 for run C91.}
  \label{fig:5791}
\end{figure}

Fig.~\ref{fig:5791} compares two runs with identical initial conditions, except 
for 
the initial $Q$ value, which was 1 for run C57 and 1.5 for run C91. We note 
that velocity dispersion does not have a big effect on spokes, except perhaps 
for a slight lowering of their intensity with increasing $Q$. 
To substantiate this impression we have measured at several time steps
the density in the gap between the two rings along a ``ring''
of the same shape as the outer ring but of smaller size, and find
that indeed that trend is present.

Fig.~\ref{fig:578689xy} compares three simulations with different target discs. 
Spokes form earlier and fade away faster in disc {\bf m1} than in 
{\bf m3}, {\bf m2} being 
intermediate. As has already been noted for rings, spokes are smoother for the 
{\bf m1} case than for the {\bf m3} 
one, where they contain multiple large clumps.

Simulation S1 shows one big massive spoke, which has a spiral rather than a 
straight line shape. This is sufficiently
clearly defined so that we can
trace back, all through the simulation, the positions of 
the particles that constitute it. At the beginning of the simulation 
they form a very tightly wound one-armed leading spiral, which, however, does 
not stand out when all particles in the disc are plotted. It unwinds with 
time and forms the spoke. This is true also for other well defined spokes in
other simulations which we tried out. As can be seen from 
Fig.~\ref{fig:s1xy}, at later 
times the particles in the spoke of run S1 do not form a tightly wound trailing 
one-armed spiral, but rather they spread out until they fill an area like a 
quadrant, whose location with respect to the disc center rotates. At even 
later times they fall more uniformly towards the center of the disc.
However the later stages of this evolution are undoubtedly influenced
by the second passage of the companion and the subsequent merging.

\subsection{Amplitude, width and eccentricity of the rings }
\label{sec:ampl}
\indent

Rings are neither always circular nor always centered on the center of the
galaxy, so, in order to find the position of the ring as a function of time, we
recenter the disc particles to the position of maximum disc density 
and split the galaxy into 12 angular sectors, each of $30^\circ$. 
We then plot radial density profiles separately for each
sector and each time. The maxima of these profiles give us the position(s) of
the ring(s) at that time and angle. The width of the ring has been 
determined by
defining as its edge on either side of the maximum either a local minimum
(mainly for the inner edge of the ring), or the radius at which the density 
drops to half the value at maximum (mainly for the outer edge). The location 
and width of the ring have been 
determined for all angular sectors and for all time 
steps for which the program could find a clear maximum on the corresponding 
radial profile. These data allow us to calculate a number of interesting 
quantities. 

\begin{figure}
  \vspace{7.00cm}
\includegraphics{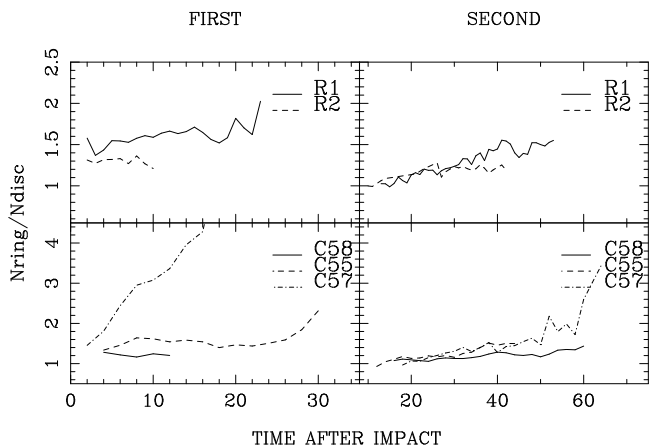}
  \caption{Amplitude of the ring as a function of time,
measured in computer units, calculated as 
described in the text. The left panels refer to the 
first rings and the right ones to the second rings. 
The upper panels compares a slow passage (R1, solid line) and a fast one 
(R2, dashed line). The lower panels compare the effects of a low mass 
companion (C58, solid line), of an intermediate mass one (C55, dashed line) 
and of a high mass one (C57, dot-dashed line).}
  \label{fig:ampliring}
\end{figure}

We have counted the number of particles within the region of the ring 
and used the ratio of this number to the number of particles in the same 
region for the unperturbed galaxy as an estimate of the amplitude at that 
position. Fig.~\ref{fig:ampliring} 
gives typical examples of the evolution of this amplitude as a function of 
time. It shows clearly that slower passages make higher amplitude first rings 
than fast ones (upper row), and a similar effect is found when comparing 
impacts by high 
mass companions to impacts with low mass ones (bottom row), as expected. The 
very large amplitudes found for the simulation which has a companion of 
double the standard mass (C57), is due to a large extent to the important 
expansions caused by the massive perturber. Thus, with our definition of the 
ring amplitude, we are comparing the ring region with a region in the outer 
parts of the unperturbed models where the density is quite low.

A similar comparison (not plotted here), now for a vertical (C59) and a 
non-vertical (C55) impact,
suggests that the impact angle does not influence much the amplitude. 
Finally 
comparing rings in target discs of different scalelength we get an indication 
that both the first and the second ring forming in the most 
extended disc have a smaller amplitude than those forming in the less extended 
one, except for the first ring in the simulations 
with single mass perturbers, where it does not seem to matter.
In general the first 
ring has a higher amplitude than the second one. Also the amplitude 
of the second ring increases with time (except for the fast encounter), 
while that of the first ring stays 
roughly constant in single mass simulations and increases considerably with 
time for impacts with heavy companions, for the reason described in the 
previous paragraph.

	With the definition given above we find for the outer ring a width of
the order of \hbox{2~--~3~kpc}, which does not vary noticeably with time.
This seems to be the opposite of what one
sees on $r=r(t)$ plots, where one gets the 
impression that the rings gets thicker
with time. The reason is that such plots 
include all azimuthal angles and, as will be
discussed below, the expansion velocity depends somewhat on the azimuth,
which gives the impression of a thickening. In general more extended discs 
have wider rings, while vertical passages create narrower ones than oblique 
passages. Similarly wider rings are generally created by slower impacts, or 
more massive companions.

\subsection{Density waves vs. material features }
\indent

As predicted theoretically (e.g., Toomre 1978, Lynds \& Toomre 1976) rings 
formed by infall of a small companion galaxy are density waves and not material 
features. We have been able to verify this by measuring what percentage of the 
particles constituting the ring at the time of its formation are still part 
of it at a given time later. Thus we find that e.g. for the first ring of 
run R1 only of the 
order of 20\% of the particles initially in the ring are still in it 6 time 
units later and hardly any 10 time units later. Similar numbers can be found 
for most other runs although larger percentages can be found for the hot and 
for the extended discs, or for impacts with high mass companions, or, in most 
cases, for the second ring.

\subsection{Kinematics }
\label{sec:kinem}
\indent

\begin{figure}
  \vspace{7.00cm}
\includegraphics{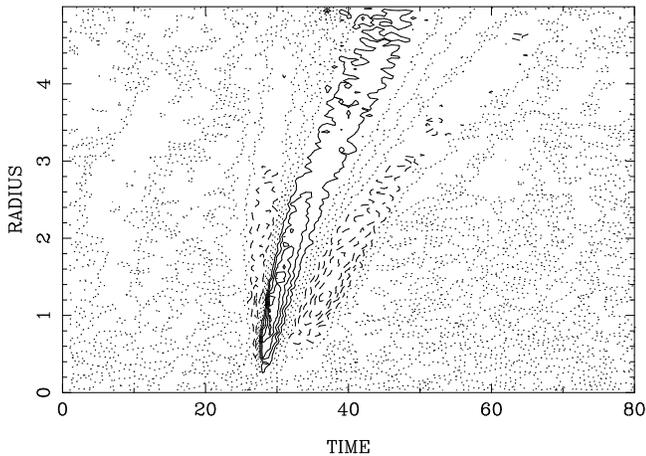}
  \caption{Isocontour plot of the radial velocity of the particles in 
           simulation R5.
           Dotted contours correspond to -10, 0 and 10 km/sec,
           solid line contours to 20, 30, 40,... km/sec and dashed ones 
           to -20, -30, -40,... km/sec.}
  \label{fig:isoradvel}
\end{figure}

Following the encounter particle orbits will first contract, then
expand, and then contract anew when
the second ring is formed. One example is seen in
Fig.~\ref{fig:isoradvel} in which we plot 
the isocontours of the radial velocity of the particles in run R5, to 
follow the evolution both as a function of time and of radius. 
Expansion is given by solid lines, 
contraction by dashed lines, and values close to 0 (positive or
negative) by dotted lines.
By comparing with the $r=r(t)$ plot
we note that the 
particles in the ring show a strong
expansion when the ring is formed, which decreases considerably
with time. Thus, during the initial stages and for a short while, expansions as
high as \hbox{40 km/sec} are not rare, 
but they rapidly fall to values of the order
of \hbox{20 km/sec} as the ring propagates outwards.
This should be compared to tangential velocities of the order
of \hbox{100 km/sec} during the 
same time interval and in the same regions. 
On either side of the ring the
particles show substantial inflow, again more important at the time of the ring
formation and decreasing with time. 
In general we find larger expansion velocities in rings
caused by more massive companions or slower impacts, while 
the extent of the target disc 
or its velocity dispersion has little 
influence on the particle velocities.

\begin{figure}
  \vspace{7.00cm}
\includegraphics{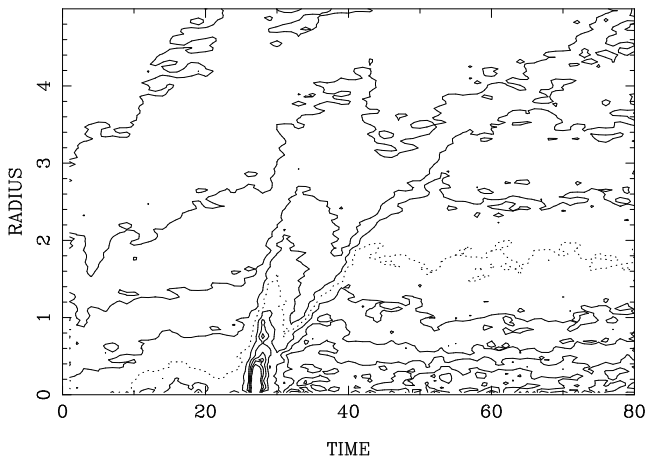}
  \caption{Isocontour of radial velocity dispersion in si\-mu\-lation
      R1. The dotted contour is at level \hbox{50 km/sec} and the other
      ones are separated by \hbox{10 km/sec}.}
  \label{fig:isodisvel}
\end{figure}

Fig.~\ref{fig:isodisvel} shows isocontours 
of the radial velocity dispersion on the $(r,t)$ plane for run R1. 
Other runs show a similar behaviour. 
The passage of the ring from any given radius produces an increase 
in the local velocity dispersion, which drops after the ring has 
passed and then increases again with the passage of the second ring. The 
other runs show a si\-mi\-lar behaviour, with the temporary rise being much higher 
for slower passages or more massive 
companions, as could be expected. This heating of the disc, occurring 
particularly after the passage of the second ring, should be responsible 
for stopping the formation of a third ring. This argument is further 
substantiated by the fact that even the formation of the second disc is 
suppressed in discs which start off relatively hot, as in simulations R5 
and R6.

\begin{figure}
  \vspace{6.00cm}
\includegraphics{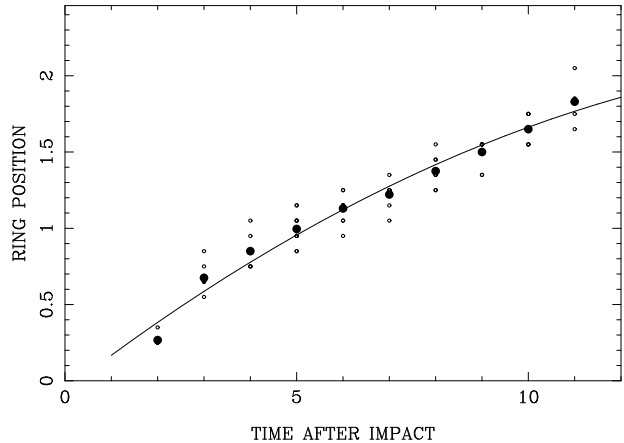}
   \caption{Position of the ring as a function of time. 
        The results are from simulation R2, and show 
        the position of the ring for different angular sectors (open 
        circles) 
        and the means over all angles (filled circles), both as a function 
         of time. The solid line is a polynomial fit to the filled circles.}
  \label{fig:ringexpvel}
\end{figure}

	To calculate the expansion velocity of the ring we have plotted the
positions of the ring measured for every angular sector as a function of time
(an example for simulation R2 is shown in Fig.~\ref{fig:ringexpvel}). This
shows that the expansion velocity depends on time and argues that it also
depends on angle, in good agreement with what one could infer from
$r=r(t)$ plots. In order to get rid of the angular dependence
we take the 
mean of all positions corresponding to the same time and then obtain the 
ring expansion velocity by fitting a second order polynomial in time
to the azimuthally averaged ring position.
We have also tried the exercise using splines, but
found that second order polynomials are more satisfactory.

\begin{figure}
  \vspace{12.00cm}
\includegraphics{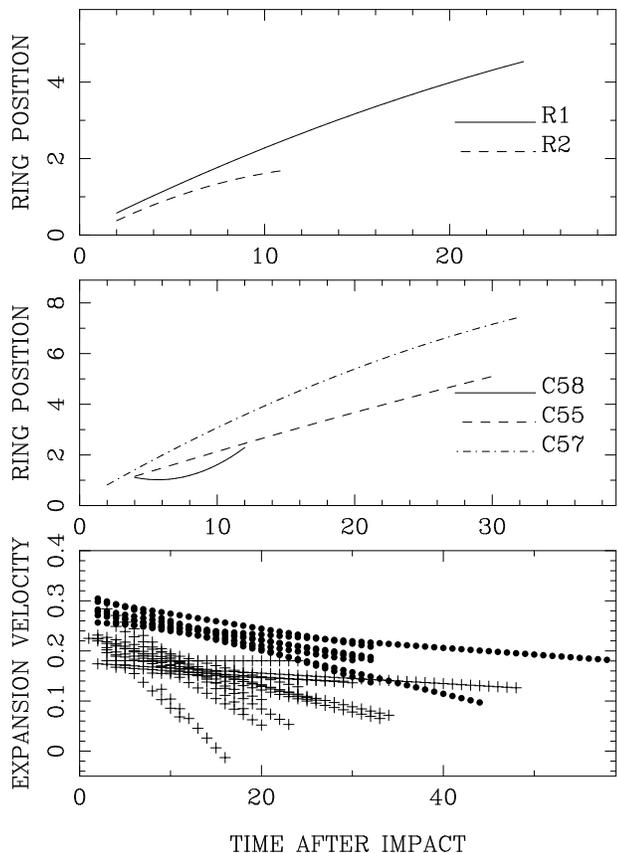}
   \caption{The upper panel gives the position 
        of the ring for a simulation with a slow passage (R1, solid line) and 
        a simulation with a fast passage (R2, dashed line) as a function of 
        time, both measured 
        in computer units. Note that the slow encounter R1
        produces a faster expanding ring than the fast
        encounter R2. In the middle panel
	we plot the position of the ring for C58 (half mass companion),
	C55 (standard mass companion) and C57 (double mass companion).
        Note that the expansion velocity of the ring depends on the 
        companion mass. The lower panel shows the expansion velocity
	of the ring in computer units. The simulations with double 
        mass companion
	are plotted with filled circles and those with standard mass
	companion with a plus sign. All three panels refer to
   	first rings.}
  \label{fig:ringexpvel14}
\end{figure}

The upper panel of
Fig.~\ref{fig:ringexpvel14} compares 
the ring position of the first ring for two runs differing by 
the velocity of the companion 
and shows that the expansion velocity of the
slow encounter (R1) is larger than that of the fast one (R2).
The middle panel of this figure compares the results for
simulations with identical initial conditions except for the
mass of the companion, and shows that larger masses produce
much faster expanding first rings. This is further illustrated
by the bottom panel, where we plot the expansion
velocity of the first ring as a function of time, this time
for all our simulations with either a standard or a double mass 
companion. It is clear that simulations
with a double mass companion (solid dots) have larger velocities
than those of simulations with standard mass companions (plus signs). 
This figure also brings out clearly the 
decrease of the expansion velocity with time.

\begin{figure}
  \vspace{12.00cm}
\includegraphics{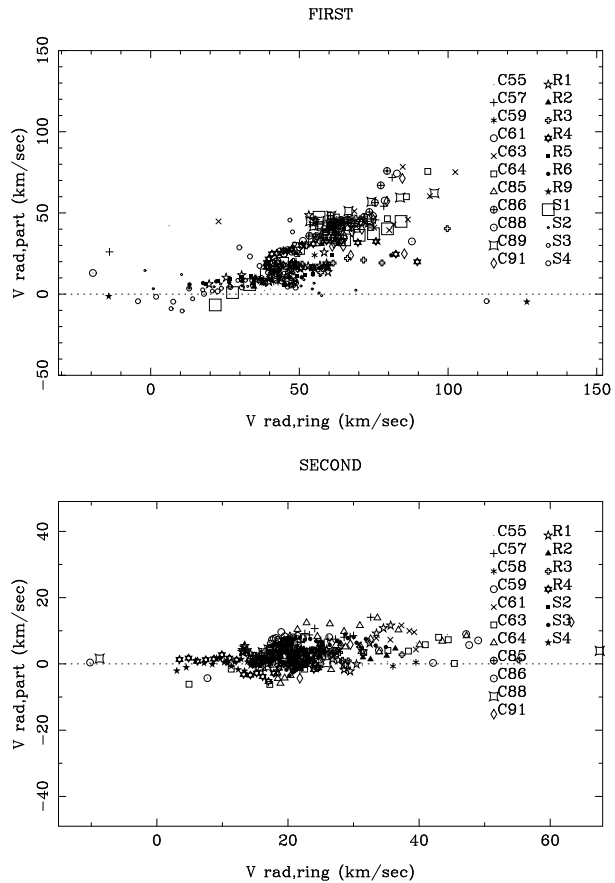}
  \caption{Expansion velocity of the ring as a function of the mean 
velocity of the particles that are in the area that defines it. There is a 
measurement for every time and every simulation. The upper panel corresponds 
to the first ring and the lower one to the second one.}
  \label{fig:partringvel}
\end{figure}

	As far as the first ring is concerned, the expansion velocity of the 
ring is larger in those cases where the expansion velocity of the particles 
is largest. This can be seen in the upper panel of 
Fig.~\ref{fig:partringvel} where we plot the 
expansion velocity of the ring at a given time, calculated as described above, 
as a function 
of the mean radial velocity of the particles in the ring 
at the same time, and 
that for all times and simulations. This trend between the two 
velocities is not true for the second ring 
(lower panel) where the expansion velocity of the ring remains small at all 
times and does not seem to depend in any clear way on the expansion velocity 
of the particles. 

\section{A barred target galaxy }
\label{sec:barred}
\indent

We also have run
three simulations in which the companion hit an initially barred 
disc. In simulations R7 and R8 we have aimed the companion at the center of 
the bar and disc, while in simulation R9 we have aimed at a point on
the bar semimajor axis.

Fig.~\ref{fig:barcentxy} shows some characteristic moments of the evolution of the barred disc 
around and after the impact of the companion in simulation R7. At time 
$t=-2$, just before the 
impact, the bar shortens somewhat and after the impact it 
gets substantially offset from 
the center of mass of the galaxy, so that it forms one side of the ring feature.
Thus at no stage of the evolution do we see a small ring with a diameter 
smaller than the bar major axis. Instead there is a rather asymmetric ring 
formed in part by the bar. This expands like the other rings discussed so far 
and, when it becomes sufficiently large, it becomes detached from the bar, 
while an arm emanating initially from one side of the bar continues the ring 
without closing it completely. Thus the result is a pseudoring, 
enclosing a bar which does not fill it completely. The evolution of 
simulation R8 is similar, so we will not show it here.

\begin{figure*}
  \vspace{17.00cm}
\includegraphics{fig18.ps}
  \caption{Evolution of simulation R7, for which the target galaxy is 
 initially barred. The bar is offcentered by the impact and forms one side of 
the expanding asymmetric ring.}
  \label{fig:barcentxy}
\end{figure*}
\begin{figure*}
  \vspace{13.00cm}
\includegraphics{fig19.ps}
  \caption{Evolution of simulation R9 following the impact.
          In this peripheral encounter the bar is almost destroyed,
          forming, in later times of the evolution, an offcentered oval
	  structure.}
  \label{fig:barperixy}
\end{figure*}

	The evolution of simulation R9 after the impact is 
shown in Fig.~\ref{fig:barperixy}. First the position of maximum density is 
displaced from the center of the bar to a point near the impact position and 
then 
the disc develops a multiarm structure. The different parts of the arms 
interact and form one very long arm, winding by nearly $270^\circ$. This 
develops into a pseudo-ring, but the density along the ring is a function of 
azimuth. The bar itself gets shorter and fatter and de\-ve\-lops into an oval, 
whose center gradually shifts towards the center of mass of the target 
galaxy. At certain stages of the evolution the result is fairly similar to 
that of a peripheral impact on a nonbarred target, except that the ``nucleus" 
is oval.
\begin{figure}
  \vspace{11.50cm}
\includegraphics{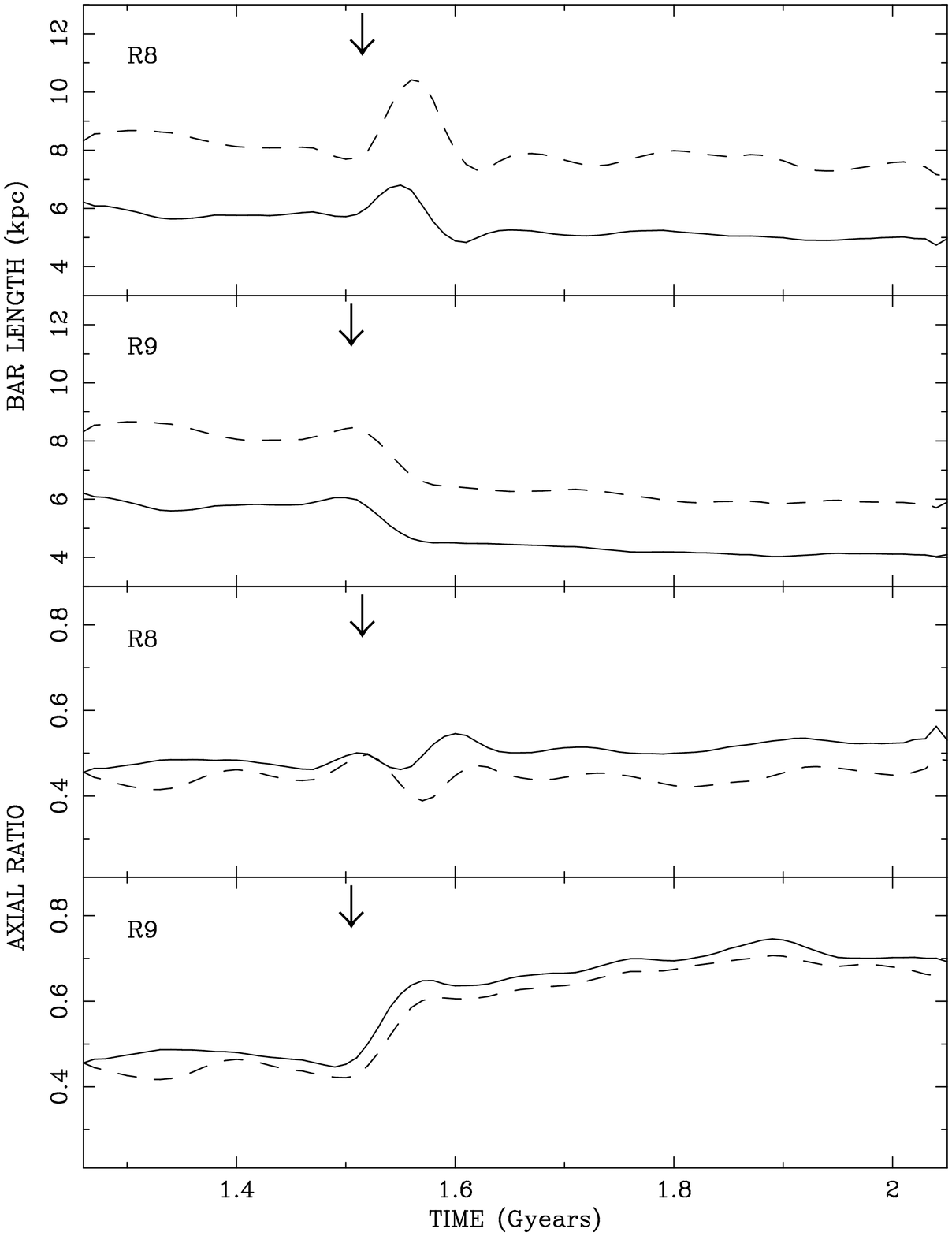}
  \caption{Bar length (upper panels) and 
         axial ratios (lower panels) as a function of time. These 
         are obtained from the isodensity contour at 0.2 (dashed line) 
         and 0.4 (solid line) times the maximum density in the
        disc at the time in question. The simulation each panel 
        refers to is given in the upper left corner. The impact 
	time is marked by an arrow.}
  \label{fig:barleneachtime}
\end{figure}
\begin{figure}
  \vspace{6.50cm}
\includegraphics{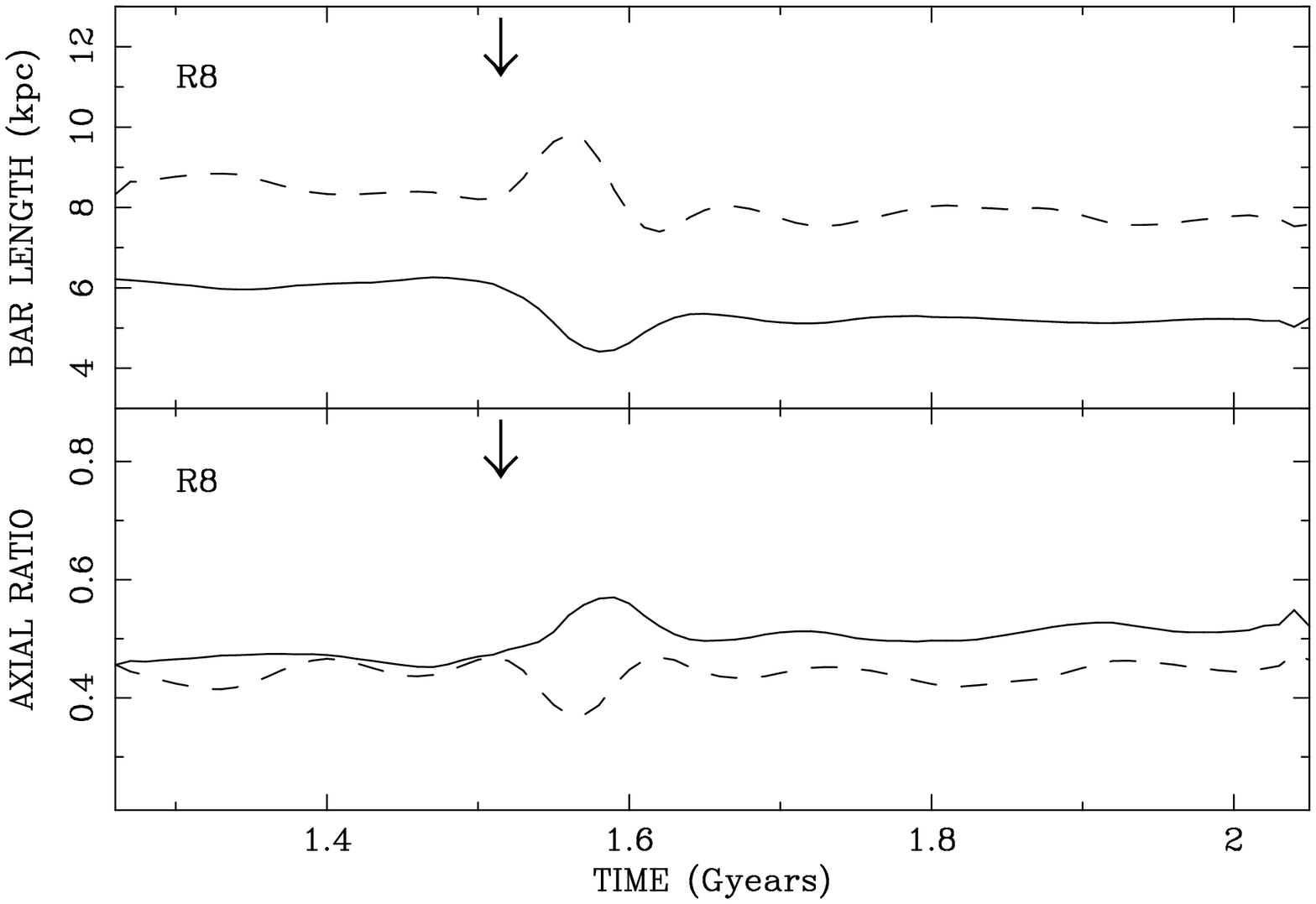}
  \caption{Bar length (upper panel) and 
	axial ratios (lower panel) as a function of time. These 
	are obtained from the isodensity contour at constant 
	projected density levels equal to 0.2 (dashed line) and 0.4
	(solid line) times the maximum density in the disc at the 
	beginning of the simulation. The impact time is marked by 
	an arrow and the simulation each panel refers to is 
	given in the upper left corner.}
  \label{fig:barlenfirsttime}
\end{figure}

	In none of the three cases was there a second ring, nor did 
transient spokes
form.

	We have calculated the distance from the position 
of the maximum density of 
the disc component, which we can define as the center of the bar, to the 
center of mass of the target galaxy at each time. 
The bar is already somewhat offset before the impact and this 
offset is substantially increased when the companion hits the disc, 
particularly so when the impact was not central. The offset lasts 
for 0.2 to 0.3 Gyears and then the bar center 
comes back to a central position.

	The pattern speed of the bar before the impact is 
in all three cases \hbox{26 km/sec/kpc}. 
Right after the impact it drops for a short time to values of the
order of 7 or \hbox{8 km/sec/kpc} for the central 
encounters and 14 for the offcenter
one. Then it rises again to 11 or 
\hbox{12 km/sec/kpc} for the central encounter and 23
for the offcenter one, values which it keeps till the end of the simulations. 

	We have made no detailed study of the shape of the bar, but we have
calculated the length of the major and minor axes at a density value
equal to 0.8, 0.6, 0.4 and 0.2 times the maximum density in the disc, and
thus obtained the axial ratios at these density levels at all times. Some of
this information is given in Fig.~\ref{fig:barleneachtime} for one central
impact (R8) and for the peripheral one (R9). The results for the second
central impact simulation are very similar, so they are not shown here. 
We have plotted
the length of the major axes of the isodensity contours at 0.4 and 0.2
times the maximum density (upper two panels), as well as the corresponding
axial ratios (lower two panels). The bar length increases considerably after
the central impact (marked by an arrow), then decreases again to a level
similar to its initial one and stays at that level till the end of the
simulation. A similar variation can be seen for the minor axis of the bar
(not plotted). Thus the bar axial ratio shows some oscillations right after
the passage of the companion then settles to a value equal to (at the 0.2
density level) or slightly higher (at the 0.4 density level) than the initial one.
On the other hand after the peripheral impact
(again marked by an arrow) the bar length drops drastically and then stays at
that level, while the minor axis increases slightly. Thus the axial ratio
increases considerably, i.e., the bar becomes more oval-like, in good
agreement with the visual impression obtained from Fig.~\ref{fig:barperixy}.
 
\begin{figure}
  \vspace{10.00cm}
\includegraphics{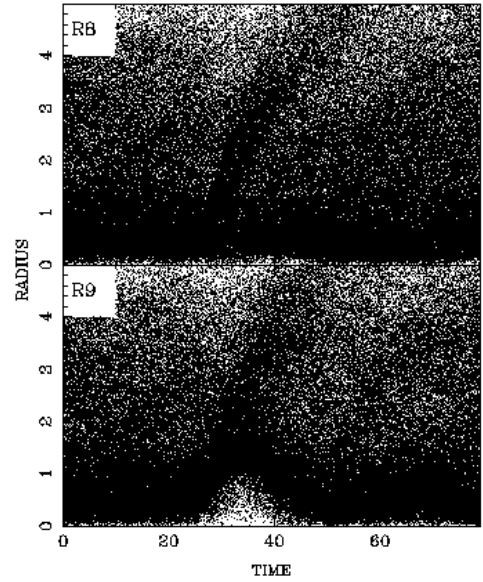}
  \caption{$r=r(t)$ for simulations R8 and R9, for which the 
	target galaxy is barred. Both radii and times are measured
	in computer units.}
  \label{fig:r=rtbar}
\end{figure}

	These diagrams are based on measurements made at density levels which
are a given fraction of the density ma\-xi\-mum and thus, since the value of the
maximum density varies with time, are not constant with time. Upon 
impact, when this is central, the maximum density increases
considerably, then drops and rises again to a value which is slightly below
it and stays constant till the end of the si\-mu\-lation. 
In the case of the peripheral impact the maximum 
density shows a small minimum after
the impact and then comes back to a value which is, within the errors, equal
the pre-impact one. Since this will affect the results of
Fig.~\ref{fig:barleneachtime} for the central impacts, we have repeated the
exercise, now using fixed density levels, those at 0.2 and 0.4 times the density
maximum in the disc at the moment the companion was introduced. The results
are given in Fig.~\ref{fig:barlenfirsttime}. They show that the results are
similar to the previous ones at the lowest density level, but that the
0.4 level (and the 0.6 and 0.8 ones, not shown here) the length of the bar
decreases after impact and then comes back to a value slightly lower then the
pre-impact one, while the axial ratio increases slightly and then drops again
to a level slightly higher than the pre-impact one. 

Fig.~\ref{fig:r=rtbar}
shows the $r = r (t)$ diagram for two of our three simulations 
with barred galaxy targets, simulation R7 not being shown since the results 
are very similar to those of R8. The bar, being a high density region, can be 
easily seen as a darker area 
at small radii. Apart from that, the formation and 
evolution of the ring show up in a way similar to that for the simulations with 
nonbarred targets discussed in the previous section. In run R9 the bar is 
severely 
offcentered for a short time and thus the darker area moves to larger radii 
leaving a low density area near the center of mass of the target galaxy. 
During the initial phases of the ring formation the bar is part of the 
ring, as was already discussed earlier, and therefore the darker area of the 
bar forms part of the ring area in the $r=r(t)$ plot of run R9. After 
$t=10$ (measured from the time of the impact)  
the ring detaches itself from the bar and continues propagating outwards, while the bar 
drifts towards the center of the target galaxy, as can be clearly followed in 
the lower panel of Fig.~\ref{fig:r=rtbar} and also on 
Fig.~\ref{fig:barperixy}.
Small displacements of the bar can also be seen 
at later times of simulation R9, and also for some time intervals in 
simulation R8, as small light areas near the $r=0$ axis.

\section{Can ring galaxies be in some cases confused with ringed galaxies? }
\label{sec:ringringed}
\indent

	Apart from the rings discussed so far, one can find in galactic discs 
rings formed by a different mechanism. 
They have been classified in three types (Buta 1986): 
outer rings, which, as their name 
indicates, are found in the outer parts of discs, inner rings, 
which are roughly of the size of 
bars or ovals, when such structures are present, and nuclear rings, 
which form
around the nucleus of the galaxy. Galaxies with such rings are known as 
ringed galaxies and good examples of this class of objects
are NGC 1291, 1433, 2217, and 7217. 

	The formation of rings in barred galaxies has been 
studied with the help 
of numerical simulations of the gas response in barred galaxy models using 
sticky particle codes, i.e., codes where gas is modelled as an ensemble of 
inelastic colliding clouds. Clouds between corotation and outer Lindblad 
resonance are pushed out by the torque exerted by the bar and form a spiral, 
which evolves to a pseudo-ring and then a ring. Such rings form in the
vicinity of resonances (Schwarz 1979, 1981, 1984, Combes \& 
Gerin 1985); the outer ring around the outer Lindblad resonance, 
the inner one around the inner ultraharmonic resonance or corotation, 
and the nuclear one around the inner Lindblad resonance. Schwarz 
associated these rings with the observed rings in disc galaxies and this was 
further corroborated by Athanassoula et al. (1982) who, analysing the ratio of 
ring diameters as given by de Vaucouleurs \& Buta (1980), found that the 
histogram of the ratio of 
outer to inner ring diameters in barred galaxies showed a sharp peak for 
values between 2 and 2.5. This is in agreement with the result of Kormendy 
(1979) that the ratio of outer ring to bar size is $2.2 \pm 0.4$, and was also 
confirmed by Buta (1995) with a sample of much higher quality, namely the 
CSRG (Catalogue of Southern Ringed Galaxies). For brevity we will often 
refer to such rings as resonance-rings.

\begin{table}
\centering
\caption{Times and distances}
\label{tab:distances}
\begin{tabular}{@{}ccccccc@{}}
& \multicolumn{3}{c|}{FIRST} & \multicolumn{3}{c|}{SECOND} \\
RUN & $T_{begin}$ & $T_{end}$ & $D_c$ & 
$T_{begin}$ & $T_{end}$ & $D_c$ \\
S1 & 0.03 & 0.18 & 0.61 & --- & --- & --- \\
S2 & 0.08 & 0.29 & 230 & 0.15 & 0.70 & 561 \\
S3 & 0.02 & 0.34 & 73 & 0.12 & 0.54 & 104 \\
S4 & 0.03 & 0.33 & 246 & 0.10 & 0.67 & 501 \\
R1 & 0.02 & 0.24 & 55 & 0.10 & 0.54 & 100 \\
R2 & 0.02 & 0.11 & 81 & 0.08 & 0.43 & 320 \\
R3 & 0.02 & 0.48 & 58 & 0.14 & 0.74 & 102 \\
R4 & 0.01 & 0.25 & 65 & 0.20 & 0.52 & 117 \\
R5 & 0.03 & 0.23 & 54 & --- & --- & --- \\
R6 & 0.05 & 0.20 & 171 & --- & --- & --- \\
R7 & 0.04 & 0.19 & 47 & --- & --- & --- \\
R8 & 0.04 & 0.19 & 47 & --- & --- & --- \\
R9 & 0.09 & 0.20 & 40 & --- & --- & ---\\
C55 & 0.04 & 0.30 & 69 & 0.12 & 0.44 & 93 \\
C57 & 0.02 & 0.36 & 60 & 0.18 & 0.74 & 87 \\
C58 & 0.04 & 0.12 & 36 & 0.10 & 0.60 & 141 \\
C59 & 0.02 & 0.22 & 55 & 0.10 & 0.74 & 146 \\
C61 & 0.02 & 0.44 & 71 & 0.12 & 0.38 & 65 \\
C63 & 0.02 & 0.34 & 59 & 0.18 & 0.44 & 69 \\
C64 & 0.02 & 0.44 & 69 & 0.18 & 0.74 & 90 \\
C85 & 0.02 & 0.48 & 106 & 0.14 & 0.74 & 152 \\
C86 & 0.02 & 0.74 & 110 & 0.24 & 0.44 & 78 \\
C88 & 0.04 & 0.26 & 66 & 0.20 & 0.52 & 118 \\
C89 & 0.04 & 0.74 & 125 & --- & --- & --- \\
C91 & 0.02 & 0.40 & 66 & 0.20 & 0.52 & 77 \\
C99 & 0.02 & 0.30 & 69 & 0.12 & 0.44 & 94 \\
\end{tabular}
\end{table}

	The situation is less clear for nonbarred ringed gala\-xies. The
driving in such cases could perhaps come from a relatively massive grand
design spiral, or from a hidden bar or oval. An alternative possibility could
be that the bar has decayed leaving behind the ring or rings, as
suggested for NGC~7217 (Athanassoula
1996, Athanassoula et al. 1996). Here we will consider whether any
galaxy with a ring formed by an infalling companion as discussed above, could
be mistaken for a ringed galaxy with resonance-rings. 

The asymmetry of rings due to peripheral and/or oblique impacts is of course 
a tell-tale sign of their origin. Also rings formed by relatively massive 
companions can
be distinguished from resonance-rings since such impacts will produce 
tell-tale spokes and, if not vertical, will severely 
distort the disc and 
form rings which, at least at their later stages of evolution, are
quite asymmetric (cf.~Fig.\ref{fig:585557xy}).
This leaves rings due to central and 
near vertical impacts with relatively less massive compa\-nions,
since their morphology can not be easily
distinguished from that of ringed galaxies (cf.~Fig.~\ref{fig:c59xy}).
Yet even in those cases one can weed out some of the interlopers by 
measuring the expansion velo\-city of the material constituting the ring, 
since, as we saw in section~\ref{sec:kinem} this is relatively high (except for
the last stages of the evolution) in rings formed by impacts.
We have also compared the ratio of the two ring radii, for ring and ringed 
galaxies, for the latter using the data by Buta (1995). Histograms
of the number of galaxies as a function of the ratio of the ring radii
peak roughly at the same position in the two cases, but there are more
ring galaxies with relative large values of the ring size ratios. 
Nevertheless, except for extreme cases, this cannot be used as a way of 
distinguishing between the two origins since individual ring galaxies
with relatively large ring ratios could always be assimilated to the
tail end of the distribution of ringed galaxies.

Is the presence of the companion the ``smoking gun" on which we can rely to 
distinguish ring galaxies from ringed ones? Table \ref{tab:distances} lists 
for all simulations discussed so far (column~1),
the time at which the 
first ring appears ($T_{begin}$; column~2) 
and disappears ($T_{end}$; column~3), 
as well as the distance from the 
companion to the center of the target galaxy at this time ($D_c$; 
column~4). 
Columns~5 to 7 contain the same information, but now for the second ring. 
In both cases time is measured in Gyears from the moment of impact and 
distance in kpc. Of course the time a ring disappears depends not only on 
personal judgement, but also on the way the data are displayed and this 
introduces nonnegligible uncertainties in our estimates. As already 
mentioned, simulation S1 starts with a 
velocity much smaller than the escape velocity 
(cf.~Tables~\ref{tab:components} 
and \ref{tab:initial}) and 
thus the compa\-nion merges with the target.
For the remaining simulations the initial velocities are considerably larger 
than the escape velocity and the companion is at a considerable distance when 
the rings disappear. The minimum distance is 36 kpc or 
3 disc radii when the first ring disappears, and 65 kpc or over 4 disc radii 
when the second ring disappears. The mean distance is 85 (159) kpc, or 17 (32) 
disc radii when the first (second) ring disappears and thus the companion may 
have gone unnoticed if the search did not extended to large distances.

We can thus summarise this section by saying that ring galaxies formed
by massive perturbers or oblique or offcentered impacts can not be 
confused with ringed galaxies. On the other hand there could perhaps be 
some confusion regarding rings resulting from low mass perturbers
with perpendicular and central impact. 
Measurements of the expansion velocities of the material constituting 
the ring should help distinguish between the two types of rings.

\section{Summary }
\label{sec:summary}
\indent

In this paper we have used N-body simulations to study the formation and
evolution of rings in disc galaxies hit by small spherical companions. 
In most simulations two transient and short-lived
rings form, although in barred or very 
hot target discs no second ring is seen.  These rings are indeed 
density waves, as 
predicted theoretically (Toomre 1978, Lynds \& Toomre 1976). The second 
ring is 
formed much after the first one, yet the two coexist for a 
considerable time span. Following the encounter particle orbits
first contract, then expand, and then contract again,
this second rebound corresponding to the formation of the second ring. 
The  amplitude of
these radial motions is a function of radius, or, equivalently, time.
In other words shortly after
the ring has formed, i.e., when it is still in the central parts of the
disc, the particles in the ring have large expansion velocities. 
At later times, when the ring has reached larger
radii, they have considerably smaller radial velocities. The passage of
the ring from any given radius increases the local velocity dispersion.

The range of companion masses considered varied between 0.02 and 0.2 times
the target mass, or, equivalently,  between 0.1 and 1 times the disc mass.
For the lowest va\-lues in this range either no ring was formed, or it was too
weak to be clearly seen. We did not investigate what the minimum mass 
necessary for ring formation is, but our si\-mu\-lations show that 0.05 of the
target mass, or 0.25 of the disc mass, are sufficient.

The amplitude, width, lifetime and expansion velocity of the first ring increase
considerably with companion mass, and so does the radial velocity of the 
particles in it. Also rings are more symmetric, narrower and 
nearer to circular in low mass encounters. Furthermore high mass impacts 
cause a substantial increase of the disc extent.
Rings formed by high mass companions are quite asymmetric during the last 
stages of their evolution. Asymmetric rings are of course formed also
by peripheral impacts.

Impacts with a relatively massive companion also make spokes, similar to
those observed in the Cartwheel galaxy. In our simulations we find the best 
spokes for $M_C/M_G~=~0.2$, although some good examples can still be found 
for $M_C/M_G=0.1$. Spokes are trailing and last only a couple of $10^8$ years.

We tried both vertical and oblique impacts, as well as fast and slow ones. 
Perpendicular impacts make rather symmetric and near-circular rings, while
oblique ones form more eccentric and broader rings. First rings formed by slow 
impacts have higher amplitudes, larger expansion velocities and longer 
lifetimes than those formed by fast impacts. Also the radial velocities of the
particles in the ring are larger.   

Discs which have relatively low surface density and extend far out 
develop multi-armed spiral structure and their rings, when they form, are
more irregular and patchy, evolving in time first to a polygon-like structure
and then breaking up into a multitude of clumps and spiral segments. Similarly
the spokes formed in such simulations are more clumpy. 

Since small companions should hit barred as well as nonbarred galaxies, we 
have also run simulations in which the target galaxy is barred. In 
all cases we noticed important displacements of the bar structure lasting 0.2 
to 0.3 Gyrs. An asymmetric pseudoring is formed in each case, one
side of which is composed by the bar. These rings expand, as the ones
formed in nonbarred galaxies, and after becoming sufficiently large they 
detach themselves from the bar. 

The pattern speed of the bar is considerably decreased during the encounter, 
and then increases again, albeit it to a value lower than the initial one. If 
we define the semimajor axis of the bar as the length at which its density 
reaches a given fraction of the maximum central density, then we note 
considerable temporary increases of this quantity after a central impact, 
followed by a decrease roughly to its initial value. On the other hand after 
a peripheral impact on the bar major axis the length  of the bar decreases 
and stays at a low value, the final product being a fat oval.

Finally we discuss whether ring galaxies could be mistaken for ringed ones,
and we argue that this could not be the case for oblique or offcentered
impacts, or impacts with relatively high mass perturbers, since these
leave tell-tale signs. Measurements of the expansion velocities of the 
material constituting 
the ring should help distinguish between the two types of rings.

\parindent=0pt
\def\rr{\par\noindent\parshape=2 0cm 14cm 1cm 13cm}
\vskip 0.7cm plus .5cm minus .5cm

\section*{Acknowledgements}

We would like to thank the referee, Ray Carlberg, for interesting comments and suggestions, Lars Hernquist for providing us with his vectorised version of the treecode, and Jean-Charles Lambert for his help with the GRAPE 
simulations and analysis software. The treecode simulations described in this 
paper were run at the C98 of the IDRIS (Institut de D\'eveloppement et des 
Resources en Informatique Scientifique, Orsay, France). 
The direct summation simulations were run locally and we would like
to thank the INSU/CNRS and the University of Aix-Marseille I for funds
to develop the necessary computing facilities. I.P. thanks the
INAOE of Mexico (Instituto Nacional de Astrof\'{\i}sica, Optica y
Electr\'onica) for a fellowship.

\bsp

\label{lastpage}

\end{document}